# In Silico Investigation of Phytoconstituents from Indian Medicinal Herb '*Tinospora cordifolia* (Giloy)' against SARS-CoV-2 (COVID-19) by Molecular Dynamics Approach

Papia Chowdhury *

Department of Physics and Materials Science and Engineering,
Jaypee Institute of Information Technology, Noida 201309, Uttar Pradesh, India.

*Corresponding author: papia.chowdhury@jiit.ac.in

## Abstract:

The recent appearance of COVID-19 virus has created a global crisis due to unavailability of any vaccine or drug that can effectively and deterministically work against it. Naturally, different possibilities (including herbal medicines having known therapeutic significance) have been explored by the scientists. The systematic scientific study (beginning with in silico study) of herbal medicines in particular and any drug in general is now possible as the structural components (proteins) of COVID-19 are already characterized. The main protease of COVID-19 virus is $M^{pro}$ or $3CL^{pro}$ which is a key CoV enzyme and an attractive drug target as it plays a pivotal role in mediating viral replication and transcription. In the present study, $3CL^{pro}$ is used to study drug:$3CL^{pro}$ interactions and thus to investigate whether all or any of the main chemical constituents of *Tinospora cordifolia* (e.g., berberine ($C_{20}H_{18}NO_4$), β-sitosterol ($C_{29}H_{50}O$), coline ($C_5H_{14}NO$), tetrahydropalmatine ($C_{21}H_{25}NO_4$) and octacosanol ($C_{28}H_{58}O$)) can be used as an anti-viral drug against SARS-CoV-2. The in silico study performed using tools of network pharmacology, molecular docking including molecular dynamics have revealed that among all considered phytochemicals in Tinospora cordifolia, berberine can regulate $3CL^{pro}$ protein's function due to its easy inhibition and thus can control viral replication. The selection of Tinospora cordifolia was motivated by the fact that the main constituents of it are known to be responsible for various antiviral activities and the treatment of jaundice, rheumatism, diabetes, etc.

## 1. Introduction:

In 2020, with the outbreak of a newly detected corona virus, the whole world is witnessing a pandemic situation which has put human life in crisis as more than 12400000 numbers of active patients with 559047 people have already died due to this virus in last couple of months (https://covid19.who.int/). The viruses, which cause illness in animals and human, coronavirus is the member of that family of viruses. In human, common cold, Middle East Respiratory Syndrome (MERS) and Severe Acute Respiratory Syndrome (SARS) are the different types of respiratory infections which these viruses can cause (Beadling & Slifka, 2004, de Wit et al., 2016). COVID-19 is the most recently declared disease by WHO caused by a new member "novel coronavirus" of this virus family (https://www.who.int/, Guo et al., 2020). Before its outbreak appeared in Wuhan, China, in late December 2019 (Wang et al., 2020; Guan et al.,2020), this new virus and disease were unknown to the human civilization. On 11 February 2020, the international committee on taxonomy of viruses declared the new novel coronavirus as "Severe Acute Respiratory Syndrome Coronavirus 2" (SARS-CoV-2) (Gorbalenya et al., 2020) (originally named 2019-nCoV). This virus is part of the Coronaviridae family of virus SARS-CoV having a genome sequences of 79.5% sequence matching (Graham et al., 2013; van der Hoek et al., 2004). The surface glycoproteins of this virus creates a crown-like appearance which can be visible under electron microscope. Crown like spike proteins are the important part of SARS-CoV-2 (Hendaus et al., 2020). It recognizes some specific human proteins. The human proteins which coats inside of the nose and the cells of lungs are mainly can interact with the spike protein of the SARS-CoV-2 virus. After interaction the two proteins bind together. In the product form, the spike protein of CoV-2 changes its shape and causes the human receptor cell to engulf the virus. After entering human body, it then replicates itself and can infect neighbouring cells and tissues by damaging mainly lung, heart, brain cells and many other organs. After few weeks of SARS-CoV-2's appearance in human society, its structural components were characterized. Results are also available on how the virus targets, interacts with human proteins and how we can fight against it (Jin et al., 2020), but till date we are searching for the answer to fight against this virus. Till now structural biologists have identified more than 160 structures of nine various kinds of SERS CoV-2 proteins by X-ray crystallography. In human, the principal target of vaccine or drug to act on is the spike protein that protrudes from the lipid shell of the virus. The infectivity of the virus would be much reduced and maybe even eliminated if and only if the human immune system are primed to recognize and counteract the spike protein of SERS CoV 2 (Khan et al., 2020). COVID-19 has now become a pandemic worldwide affecting mostly all countries globally. It is an infectious disease with a high potential for transmission for human to human close contacts through the respiratory droplets (such as coughing) and by fomites that can propagate through air at a minimum distance of 1 meter (Rothe et al., 2020). Research works performed till now suggests that maintaining a distance of more than 1 meter between two individuals which is termed as ‘social distancing’ or ‘physical distancing’ along with proper hand-hygiene reduces the chance of being infected by COVID-19. Though there are many predictions about the airborne transmission of this disease, but until now no scientifically valid evidence is available (Doremalen et al., 2020). Fever, dry cough, tiredness and less working of sensing organs like testless feeling of food, less sensing of smell etc are the most common symptoms of COVID-19. These symptoms are appeared usually in mild form and begin gradually afterwards. Few infected people even become asymptomatic or only have very mild symptoms. The main point of concern is that till now we don’t have any specific therapies for COVID-19. Research results regarding the treatment of COVID-19 are not very productive also (Weiss & Murdoch, 2020). As per current data of July 20th, 2020, over 12401262 cases of COVID-19 active patients have been confirmed worldwide and over 559047 death cases have been reported so far. Due to nonavailability of any vaccine or medicine, the supportive care and non-specific treatment to ameliorate the symptoms of the patient are the only options currently to treat COVID-19 (Salata et al., 2020). However, extensive research works leading to clinical trials of both western and traditional medicines are going on all over the world to develop vaccines and medicines to prevent and cure CoV-2 infections (Jin et al., 2020, Bhardwaj et al, 2020, Rameez et al 2020, Enmozhi et al 2020). To combat COVID-19, some preliminary research observations have indicated that the combinations of some clinically applied anti-malarial drugs (e.g., Chloroquine, Hydroxychloroquine) (Chang et al., 2020, Khan et al., 2020) and anti-HIV vaccines can be used to treat COVID-19. Also some conventional FDA approved drugs (Remdesivir, Nelfinavir, Paritaprevir, Raltegravir, Praziquantel, etc.) are being tried as the potential drug against this CoV-2 and found with certain curative effect *in vitro* (Hendaus et al., 2020, Khan, Jha, et al., 2020, Wang et al., 2020). But till no none of the clinically applied drug/vaccine response is found to be very encouraging. Also toxicity of the currently applied drugs remain an inevitable issue causing serious adverse effects to the patients (Grein et al., 2020, Wang et al., 2020; Kaisari & Borruat, 2020). In short, no specific therapy, medicine or vacine for the effective treatment of COVID-19 has yet been reported. This unavailability of cure to this disease motivated us to investigate the possibility of inhibition of CoV-2 by some phytoconstituents available from some Indian medicinal plants. It is a well-known fact that in the treatment of pandemic and endemic diseases, Medicinal herb extracts have accumulated thousands of years’ experience (Jee et al., 2018; Kumar et al., 2018). To get a valid answer to act against COVID-19, various complementary and alternative vaccine, medicine and treatment methods are also needed for the managements of patients with SARS-CoV-2 infection. So research works with different phytoconstituents from various traditional medicinal herbs is certainly worth of investigation (Khaerunnisa et al 2020) in this present emergency situation. To deal with this emergency for the present work, we have chosen an Indian medicinal plant: Guduchi Pippali (Giloy) or *Tinospora cordifolia* (Figure 1). *Tinospora cordifolia* is the member of the family of Menispermaceae. It is usually found in mostly Asian counties like India, Myanmar, Sri Lanka, and China. The plant is commonly used as a main component in many traditional Ayurvedic medicines. It is used as a parent medicine for the therapies against several common diseases like jaundice, rheumatism, urinary disorder, skin diseases, diabetes, anemia, inflammation, allergic condition, etc. (Sonkamble et al, 2015). For the treatment of helminthiasis, heart diseases, leprosy, rheumatoid arthritis, etc the stem of the plant is very useful. It also supports the immune system by increasing the body's resistance to various infections, supports standard white blood cell structure, function, and levels (Sharma et al 2019). All of the above mentioned pharmacological actions of *Tinospora cordifolia* originates from its chemical constituents of different parts like leaves, stem, root, flower, seed etc. The various categories of chemical constituents of different classes such as alkaloids, glycosides, steroids, phenolics, aliphatic compounds, polysaccharides essential oils, a mixture of fatty acids, and polysaccharides are present in different part of the plant body, including root and stem. Phytochemistry of all these are well documented in the literature (Sharma et al 2019).

| | |
|---|---|
| 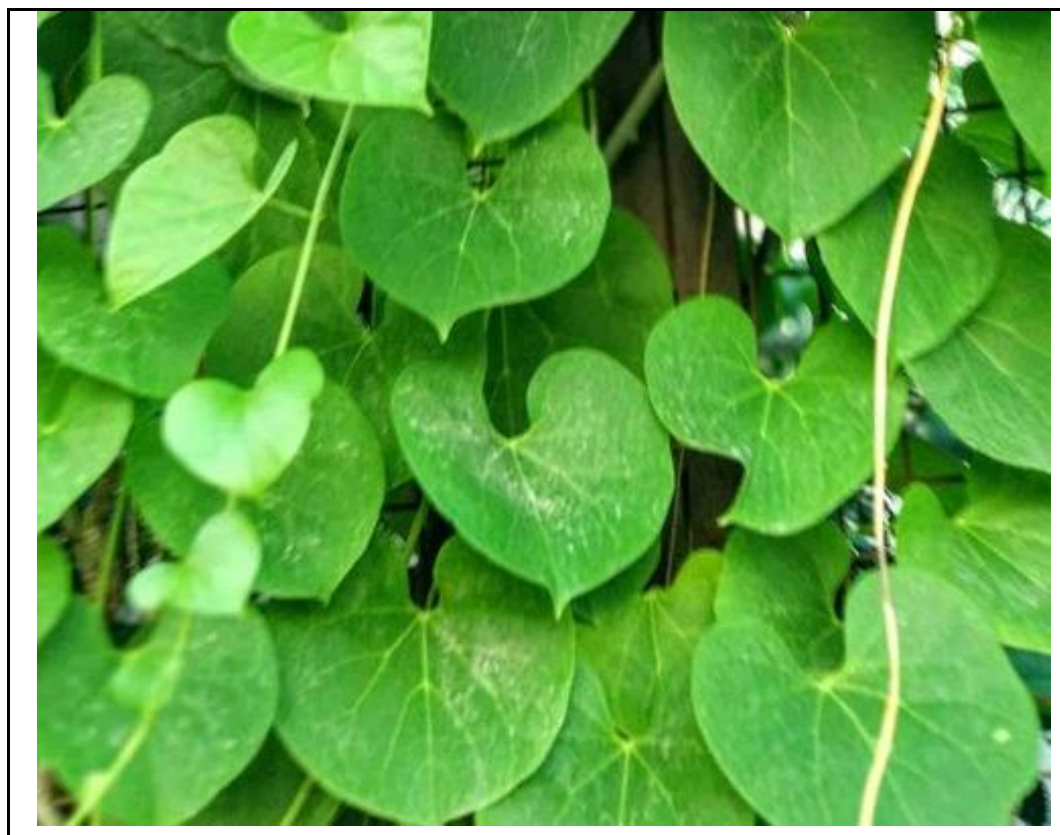 | 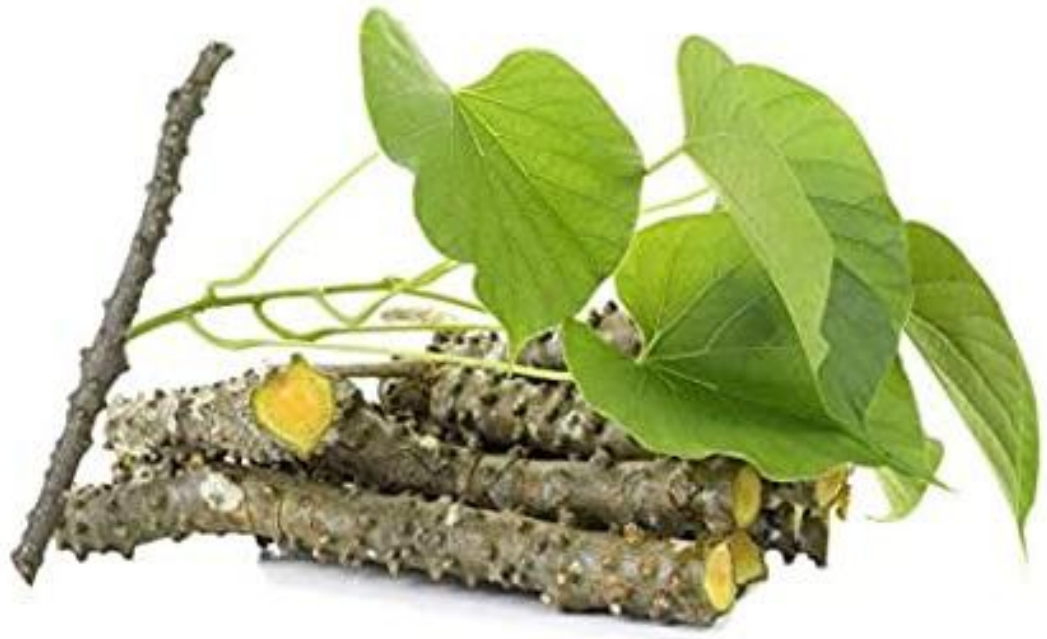 |
| Leaf of *Tinospora cordifolia* | Stem of *Tinospora cordifolia* |

**Figure 1:** Morphology of *Tinospora cordifolia*

Estimation of pharmacological properties of small molecules when it interacts with target protein is as a crucial step for the drug discovery. We have selected a few small sized phytoconstituents extracted compounds from *Tinospora cordifolia* (Figure 2) (https://www.drugbank.ca/drugs). These phytoconstituents were selected after applying some proper virtual screening by evaluating their drug-likeness, pharmacokinetics and lipophilicity properties which are set of guidelines for identification of potential drug compounds. After virtual screening, we have evaluated their potential inhibition properties against CoV-2 main protease by the mechanism of molecular docking. After studying proper binding affinity of proposed drugs against the target receptor by the Molecular docking, finally we have verified the drug inhibition with target receptor by different interaction energies like hydrogen bonding, hydrophobic, coulombic etc by MD simulation approach to validate the applicability of proposed drug molecule as suggested by Anuj Kumar et al 2020. *In vivo* estimation of drug molecules is time consuming and expensive so to deal with the present COVID crisis, *in silico* methods have become inevitable as we need an urgent solution of SARS-CoV-2 infection.

| Compound Name with symbol | Drug bank ID | Chemical Structure | |
|---|---|---|---|
| | | Two dimensional | Three dimensional |
| berberine ($C_{20}H_{18}NO_4$) | DB04115 | 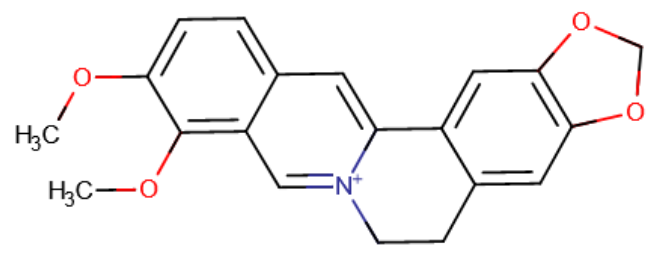 | 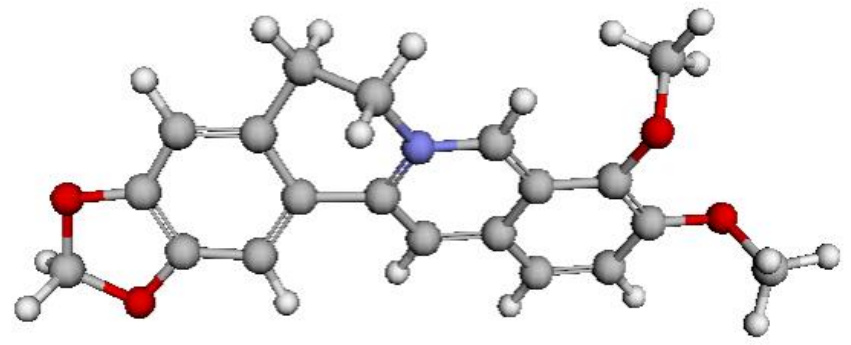 |
| choline ($C_5H_{14}NO$) | DB00122 | 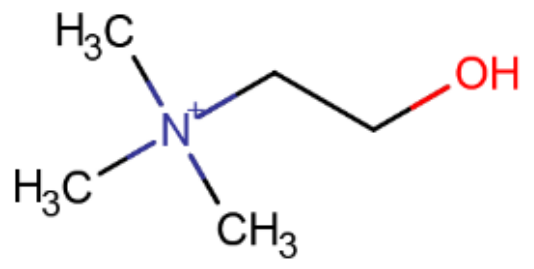 | 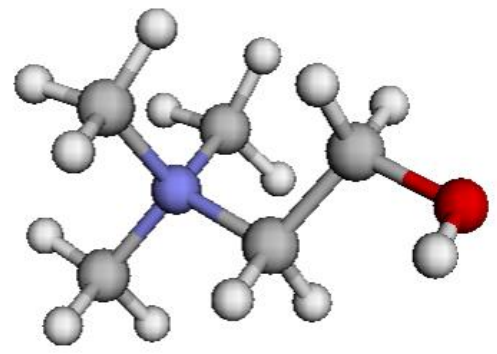 |
| β-sitosterol ($C_{29}H_{50}O$) | DB14038 | 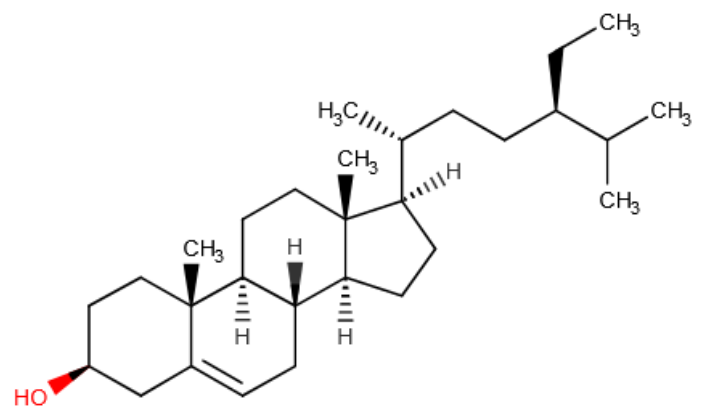 | 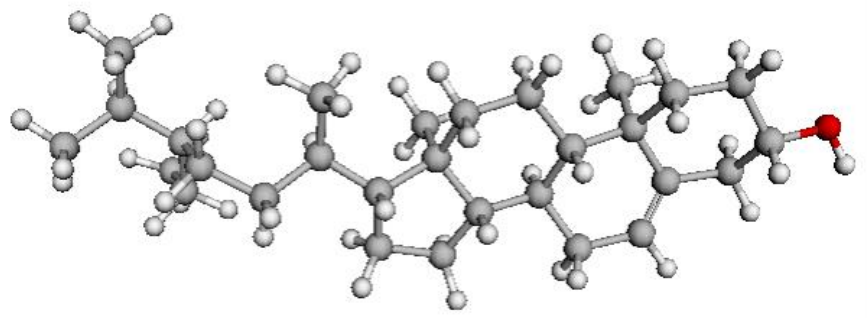 |
| tetrahydropalmatine ($C_{21}H_{25}NO_4$) | DB12093 | 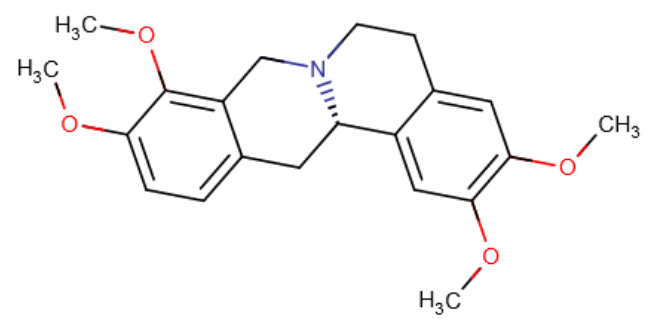 | 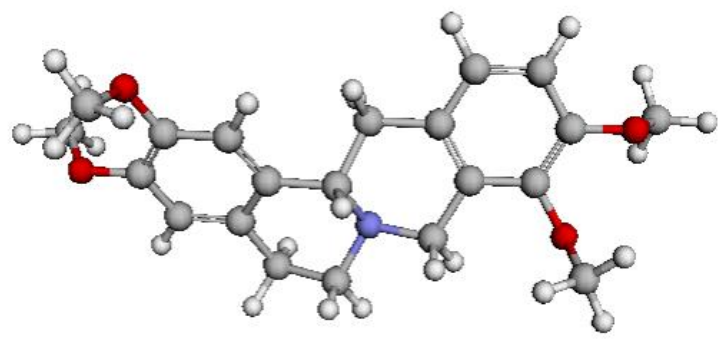 |
| octacosanol ($C_{28}H_{58}O$) | DB11220 | 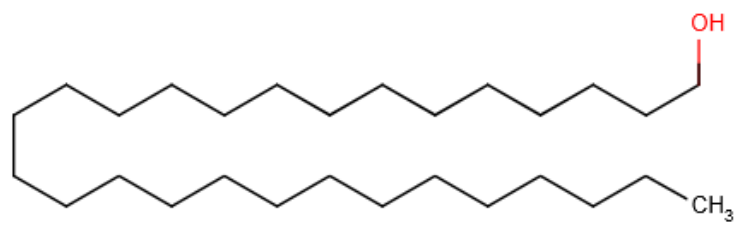 | 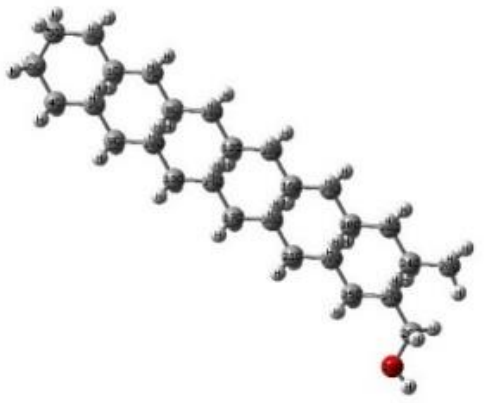 # |

**Figure 2:** Different phytoconstituents extracted compounds from Guduchi Pippali (Giloy) or *Tinospora cordifolia* as ligand molecules. [# from Gausswiew, rest from Drugbank].

Important pharmacological information of various chemical constituents have indicated the research community to explore computational molecular docking and MD methods for investigating the drug-protein interactions. Variety of simulation methods and databases have yet been used for the *in-silico* prediction of target drugs (Feixiong et al 2013, Boopathi et al 2020). Many groups are presently working on different quantum mechanical simulation techniques to search varieties of interacted complex configurations of different organic and inorganic systems and their interactions with different environments (Chowdhury et al 2017). We have earlier studied the environmental effects on different organic probe molecules in view of their applications in various biomedical field (Chowdhury et al 2014). In the presented study, we have reported some encouraging responses we obtained in terms of inhibition potentials from some of the tested phytoconstituents of current probe system: *Tinospora cordifolia*. We strongly consider that the outputs of the present work will lead to some important insights into the development of alternative drugs for COVID-19.

## 2. Materials and Methods:

### 2.1. Potential Target Protein Structures for SARS-CoV-2

Coronaviruses have positive-sense single-stranded RNA. Coronavirus SARS-CoV-2, responsible for COVID -19 is member of β-coronavirus genus (Fehr & Perlman, 2015). CoV-2 contains genome size of ~30 kilobases. It encodes for various non-structural and structural proteins. The structural protein components of SARS-CoV-2 comprise: envelope (E), membrane (M), spike (S) and nucleocapsid (N) (Gorden et al., 2020; Woo et al., 2020). Structure of SARS-CoV-2 has been identified quickly after it first appearance and its genomic sequence is already available (Wu et al., 2020) for us. As the virus has been discovered very recently so very few available immunological information (immunogenic epitopes eliciting antibody or T cell responses) about the virus are available until now. It is a fact that until now, no available therapeutics or approved and tested vaccines/medicine are there to act against any human-infecting coronaviruses. The newly invented SARS-CoV-2 is very closely related to SARS-CoV structures. So utilization of this known protein structure can be applicable for quick discovery of drugs against this newly appeared virus (Jiang et al 2020). It is reported very recently that CoV-2 enters into host cells by the spike (S) glycoprotein and forms homo trimers protruding from the viral surface (Jin et al., 2020). This spike protein interacts strongly with the human ACE2 (angiotensin-converting enzyme 2) receptor (Velavan & Meyer, 2020). After entering the host cell, SARS-CoV-2 replicates itself through some cyclic processes. First it translates its genomic RNA (gRNA). Proteolysis of the translated polyprotein takes place with viral 3C-like proteinase. After that replication of gRNA takes place. The viral replication complex appears which consists of RNA dependent RNA polymerase (RdRp), helicase, 30-to-50 exonuclease, endoRNAse, and 20-O-ribose methyltransferase. Lastly the assembly of viral components takes place. The proteins (S) which are associated with replication, are the primary targets of post-entry treatment drugs or vaccine design which can suppress viral replication/infection or neutralizing antibodies (Abs) upon infection (Gao et al., 2020; Gupta et al., 2020; Khan et al., 2020; Sarma et al., 2020). SARS-CoV-2 3C like proteinase is already predicted to bind with different FDA approved antiviral commercially available drugs like atazanavir, lopinavir, ritonavir, remdesivir, efavirenz and some other antiviral drugs which have a predicted affinity with clear efficacy value of $K_d$ > 100 nM potency (Hendaus et al., 2020, Khan, Jha, et al., 2020). Prediction suggests that viral proteinase targeting drugs should act very efficiently on the viral replication process. In support of the prediction, it was reported by the docking study of some HIV proteinase inhibitors of the CoV proteinase that lopinavir, atazanavir and ritonavir may inhibit the CoV proteinase (Dayer et al 2017). Like lopinavir, atazanavir and ritonavir there are some other case studies also with other inhibitor drugs like hydroxyxhloroquine, remidisivir etc., but until now there is no real evidence about whether these drugs can act efficiently as predicted against COVID-19. Traditional drug discovery usually takes years of research and trial. Now the whole world is dealing with an emergency situation with an urgent need of required drug against this virus to save the infected lives. An alternative approach to complement the existing procedure and to combat this problem can be to use the computer aided drug design (CADD) or computer-assisted structure-based drug design (SBDD). For the present work, some particular type of SBDD approaches have been implemented. Specifically, we tried here the *in silico* docking model and MD simulation approaches to search for medication of COVID-19 by using two most variable viral proteins (I, II) SARS-CoV-2 protease enzyme $M^{pro}$ or $3CL^{pro}$ as the receptor (Figure 3a,b).

<table>
<tr><td>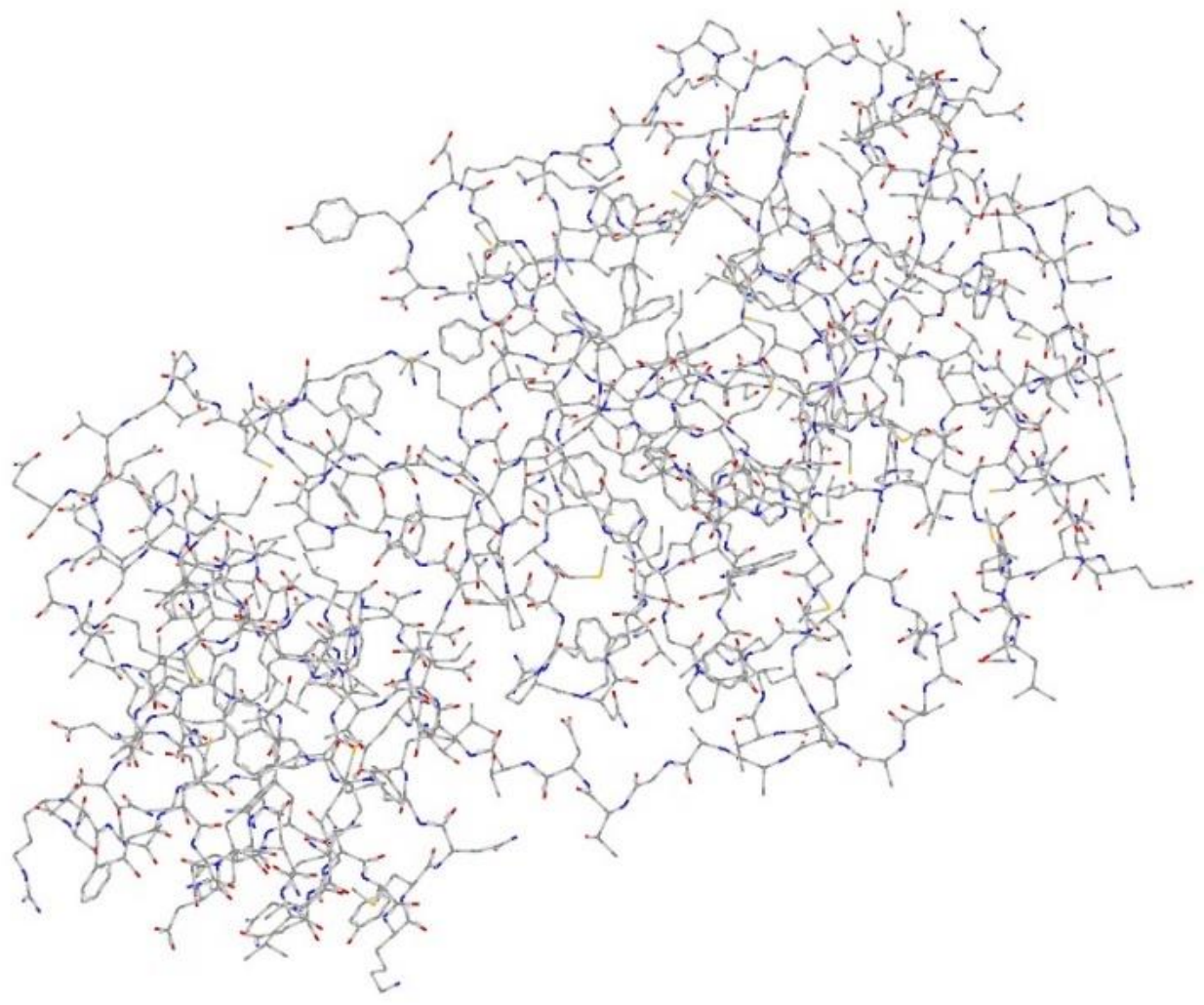</td><td>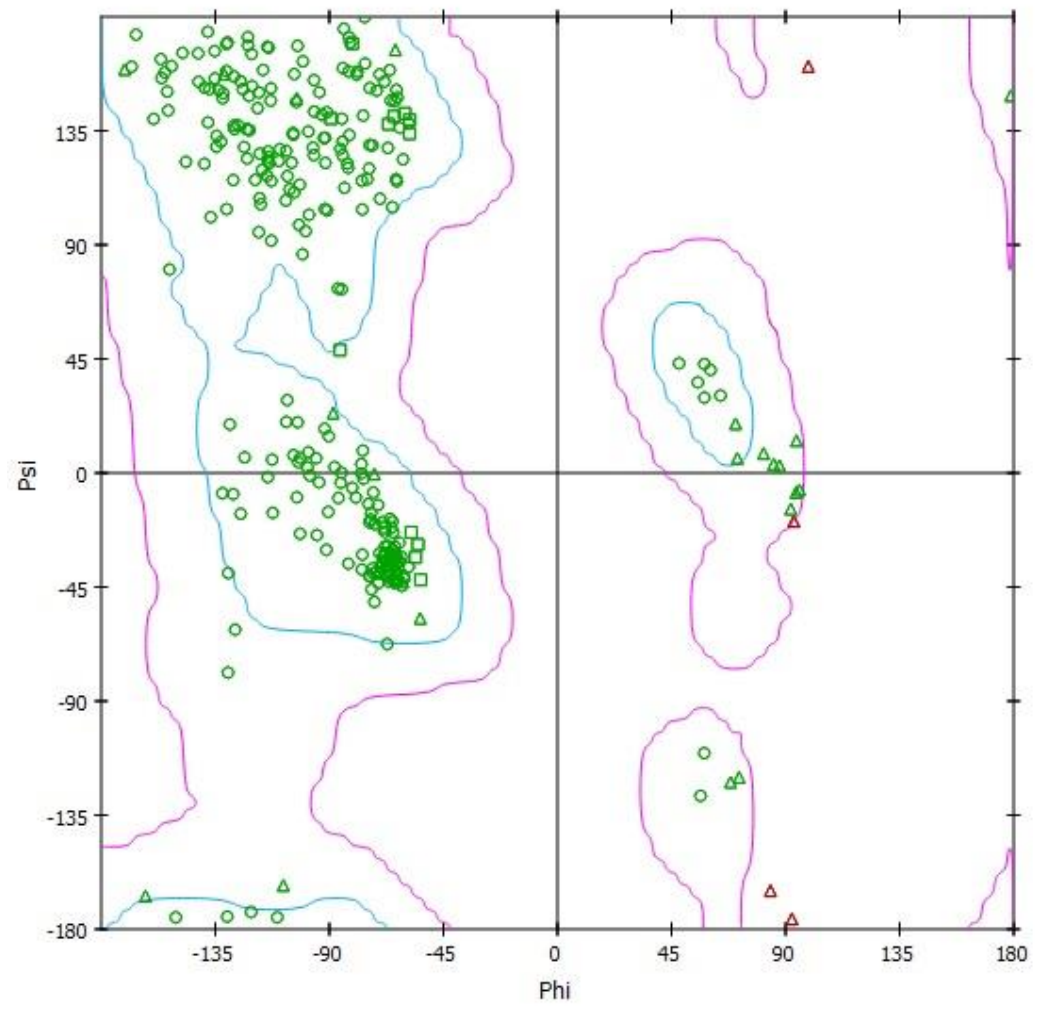
</td></tr>
<tr><td>(a)Target protein I: $3CL^{pro}$ (PDBID: 6LU7) without added hydrogen and inhibitor</td><td>Ramchandran plot for 6LU7</td></tr>
<tr><td>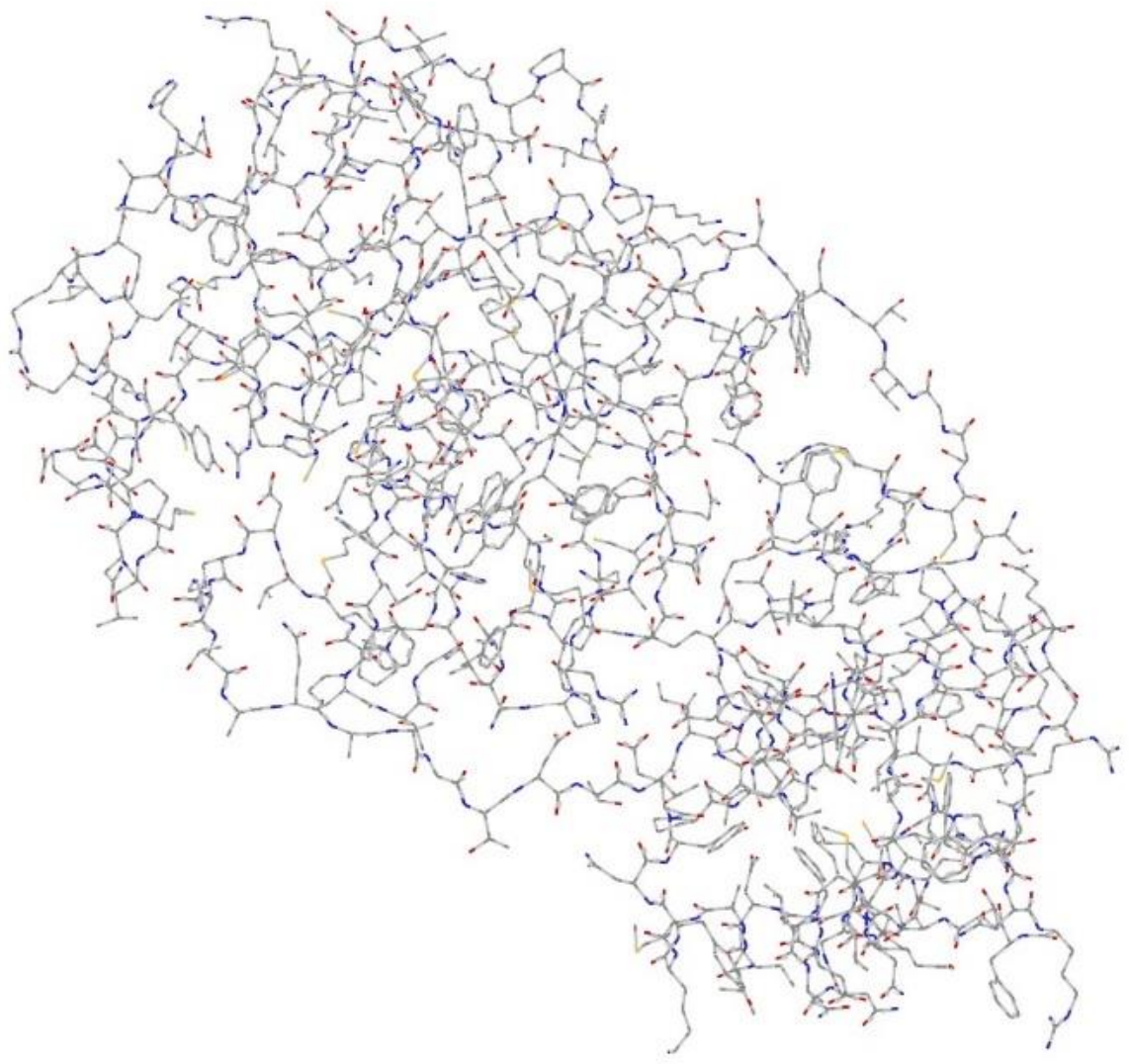</td><td>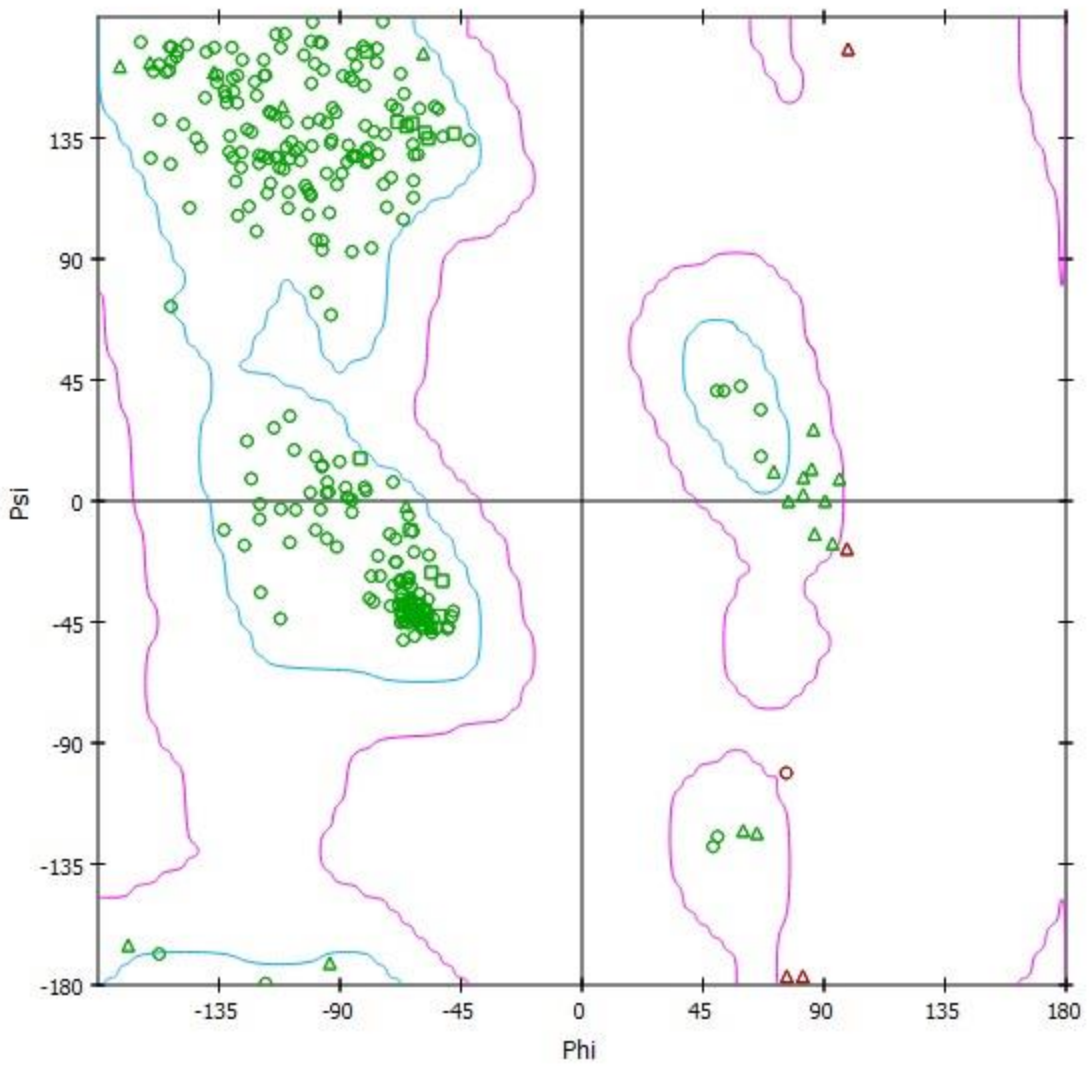
</td></tr>
<tr><td>(b) Target protein II: $3CL^{pro}$ (PDBID: 6W63) without added hydrogen and inhibitor</td><td>Ramchandran plot for 6W63</td></tr>
</table>

**Figure 3a,b:** (a).Target variable viral proteins (I, II) SARS-CoV-2 protease enzyme $M^{pro}$ or $3CL^{pro}$ as the receptor. (b) Ramachandran plot for both the receptor protein.

3CL$^{pro}$ is a key to SERS-CoV enzyme. CoV 3CL$^{pro}$ is responsible for the maturation of itself and the subsequent maturation of the replicate polyproteins. This protein has a conserved catalytic domain. The protein can control functions of the virus like: replication and transcription processes. Such important functionalities of 3CL$^{pro}$ makes it an ideal target for drug development (Anand et al 2003). Also due to its viral life cycle and the absence of closely related homologues in human, the 3CL$^{pro}$ protease has become most attractive receptor protein for antiviral drug design. Recently 3D structure of one of the 3CL$^{pro}$ like protease protein is reported by X-Ray crystallographic data (PDBID: 6LU7) (Jin et al., 2020). Protease proteins are essential to the transmission and virulence of the virus. For an effective therapy, by inhibiting these protease proteins with proper drug inhibitor, the severity of the infection can definitely be reduced. We have tried to check the inhibiting and binding possibilities of two of these natural 3CL$^{pro}$ like protease protein substrates in order to find the possibilities of effective drug design for COVID-19 in the present work.

## 2.2 Protein Receptors Preparation

In this work we have used two CoV-2 3CL$^{pro}$ main proteases as target of the potential drug molecules. The proteases used are: protease in complex with an inhibitor N3 (PDBID: 6LU7) (Jin et al., 2020, Zhang et al 2020) referred to as I and the other one is protease bound to potent broad-spectrum non-covalent inhibitor X77 (PDBID: 6W63) referred to as II (Mesecar et al 2020). The 3D structures of both were downloaded from the Protein Data Bank (https://www.rcsb.org/) and are shown in (Figure 3a,b) without added water and hydrogen. All the available required properties of these proteins are described in Table SD1. At the beginning of the present investigation, all protein structures have been cleaned by removing the existing lead components. Also the existing water molecules and ions have been removed from protein structures. After cleaning, Gasteiger charges of protein structures have been calculated and procedure of inclusion of polar hydrogens have been performed. Then subsequently the merging of non-polar and rotatable bonds were defined. All of the above mentioned steps have been performed by Auto Dock and MG Tools of Auto DockVina software (Trott et al 2010). Lastly the inbuilt ligands were removed from the protein molecules using Discovery studio 2020 (Dassault Systèmes BIOVIA) and the cleaned and final proteins structures were saved in PDB format. Active amino acid residues which will be involved in mediating enzyme activity for protein-ligand interaction were predicted for each target protein using Ramachandran plot (Figure 3a,b) by visualizing the dihedral angles $\psi$ against $\varphi$ of amino acid residues. This two-dimensional plot shows the allowed and disfavoured values of $\psi$ and $\varphi$. In the present work for our target 3CL$^{pro}$ proteins, each plot specifies localization on the A chain residues. This localization determines the quality of the protein structure. Good quality reflects the efficient and accurate docking possibilities. In the present work, the Ramachandran plot for both the target proteins show good localization of residues which are the indication of good probable docking results.

## 2.3 Ligand drug molecules Preparations

For drug discovery, estimating required *in vivo* and *in silico* pharmacological properties of proposed small sized molecules are reflected as a very essential step. *In vivo* estimation is expensive and time consuming and so in present scenario *in silico* methods have become certain. For *in silico* estimation the process is known as virtual screening. In virtual screening different Drug-likeness rules (a set of guidelines for the determination of structural properties of proposed drug compounds) are used for preliminary screening of required drug-like properties of a molecule. Some mostly industry applicable drug-likeness rules are: Lipinski's rule, MDDR-like rule, Veber's rule, Ghose filter, Egan rule, Muegge rule etc. Out of these different methods (rules), one of the dominant method is Lipinski's rule of five (Ro5) which is known as "a rule of thumb". It is used to evaluate if a proposed chemical compound possesses certain required pharmacological properties which would make the compound a potential candidate for orally active drug in humans (Lipinski et al 2004). Candidate drugs that imitate Ro5, usually have an increased chance of reaching the market since they tend to have lower attrition rates during clinical trials. Ro5, with its filters (Molecular weight ≤500, H-bond donors≤5, H-bond acceptor≤10, MLOGP≤4.15, molar refractivity should be between 30-140) helps in distinguishing between non-drug like and drug like molecules which helps to circumvent preclinical and clinical failures by early preclinical development of drugs. We have done the virtual screening of our proposed drug molecules by applying most of the above mentioned drug-likeness rules (Veber et al 2002) including some other screening properties like polar surface area (Daina et al., 2017), physicochemical properties, pharmacokinetics, lipophilicity, synthetic accessibility etc (Table 1). ADME (Adsorption, Distribution, Metabolism and Excreation) is important technique to analyze the pharmacodynamics of the proposed drug molecule. SWISS-ADME software (https://www.swissadme.ch) helps us to identify proposed drug molecule by applying different virtual screening methods. Different components of lipophilicity (iLOGP, WLOGP, XLOGP3, MLOGP, Log Po/w), water solubility (Log S (SILICOS-IT)), pharmacokinetics (GI absorption, BBB permeant, P-gp substrate, Log ($K_p$)) also help us to preliminary test the suitable drug molecule. For choosing the proposed drug molecules through virtual screening, sometimes we found some of them have violated some selection rules like Ro5, Veber etc. In drug industry today there are several important drugs available in market who violate some likeness rules. Some very popular drugs like dabigitran etexilate, bromocriptine mesylate, olmesartan medoxomil, fosinapril, and reserpine etc revealed two Ro5 rule violations (Giacomini et al 2010). Some very popular tyrosine kinase inhibitors (e.g. lapatinib and nilotinib) and HIV protease inhibitors (e.g., lopinavir, atazanavir, nelfinavir) do not follow the Ro5 criteria. Atazanavir exhibits three Ro5 violations. Keeping all these points in mind we have considered some natural compounds as our potential inhibitors for $3CL^{pro}$ protease although some of them violate one or two criteria of Ro5 but validate other criteria. After thoroughly testing the proposed phytoconstituents by all mentioned virtual screening of drug-likeness properties we have tested the drug binding possibilities with target $3CL^{pro}$ main proteases by molecular docking and further MD simulation mechanisms. Specifically, based on the positive responses from virtual screening, we have selected five phytoconstituents from *Tinospora cordifolia* as potential ligand drugs: berberine ($C_{20}H_{18}NO_4$), choline ($C_5H_{14}NO$) and tetrahydropalmatine ($C_{21}H_{25}NO_4$) from the group of alkaloids, β-sitosterol ($C_{29}H_{50}O$) from steroids, octacosanol ($C_{28}H_{58}O$) from aliphatic group. Details of the structure of these molecules were downloaded from Drug Bank in pdb format and are described in Figure 2 and Table 1(a-e) along with their various chemical, physical, pharmakokinetics and drug likeness properties obtained from SWISS ADME. All 2D and 3D view of the proposed ligands were obtained from Drugbank website. For octacosanol we have used Gaussian supported Gaussview to draw 3D structure (Figure 2). All of our proposed drug molecules have molecular weight less than 500 g /mol. All of them have TPSA values less than 50 $Å^2$. All drug molecules have H-bond donors≤5, H-bond acceptor≤10. All of the proposed drug molecules have synthetic accessibility count between less than 10 so that they can be synthesized easily. Though β-sitosterol and octacosanol have violated some of the drug likeness properties, still we have consider these phytoconstituents as potential inhibitors due to their availability in drug industry. For example, octacosanol is used as a probable therapeutic agent for treating Parkinson's disease. It also possess antiaggregatory properties and cholesterol-lowering effects. It is used as a component of Ginsamin Power as a tablet from Biogrand Co., Ltd (https://www.drugbank.ca/). It has passed the clinical trial also for the treatment of high Cholesterol condition (Taylor et al 2003). For Molecular docking study, the ligand file is required in pbdqt format. These ligand drug molecules have been saved in pdbqt format by Auto Dock Tools 1.5.6 (Michel et al 1999).

## 2.4. Molecular docking and Visualization

Molecular docking predicts the potential drug-target interactions by energy minimization and binding energy calculations. Interaction between small molecules (ligand) and protein receptor (may be an enzyme) creates the possibility of inhibition of ligand towards receptor enzyme. So output of molecular docking can demonstrate the feasibility of any biochemical reaction. It can also predict an optimized orientation of ligand towards the target receptor. By different binding modes of ligand with target receptor, docking result may provide a raw material for the rational drug designing. In combination with scoring function, Molecular docking method is used for finding more potent, selective and efficient drug candidates. The docking algorithm performs through some cyclical processes. The cyclic process by specific scoring functions helps to identify the suitable ligand conformation. After reaching to a minimum energy, the process converges to a solution (Yuriev et al 2011). Docking mechanism uses inhibitor drug's binding properties with nucleic acid of target protein. The binding property establishes a correlation between drug's molecular structure and cytotoxicity. Molecular docking can also predict whether the ligand/drug is docked with the receptor protein/DNA or not. Docking algorithm gives output in the form of quantitative predictions of binding energetics by providing the rankings of docked compounds. The ranking is based on the binding affinity of ligand: receptor complexes. The best confirmation of the ligand: receptor complex is that or those which has/have lower binding energy. After getting positive prediction, further experimental procedures are applied to the development of new drug molecule. Docking-based studies on the suggested inhibitors onto the protease of CoV-2 through Receptor: ligand docking analysis was performed using AutoDock Vina (Trott et al 2010). The macromolecule (protein) file is saved in pdbqt format by Auto Dock Tools (Trott et al 2010) and ready to be used for docking. Similarly the ligand molecule has been saved in pdbqt format by Auto Dock Tools. For AutoDock Vina algorithm, for setting up the configuration file following parameters were used: (i) count of binding modes- 9; (ii) exhaustiveness - 8 and (iii) applied maximum energy difference- 3 kcal/mol. Ligand centered maps were generated and Gridbox center was set to coordinate x y, and z of residue position of the target protein respectively.

Out of all the possible poses (optimized ligand: protein complex structure) suggested by simulation according to the binding modes, the pose showing maximum hydrogen bonds and minimum binding affinity(kcal/mol), were chosen as the best ligand: protein complex structure formed by ligand-receptor interaction. For choosing the best possible ligand: protein complex structure we have taken the help of root means square deviation (RMSD) method. For RMSD evaluation only movable heavy atoms (ligand) were considered to identify the best mode. Two variants of RMSD: RMSD/lb (lower bound) and RMSD/ub (upper bound) were used here for the whole simulation. The two variants differ in how the atoms are matched in the distance calculations. For the distance for upper bound is considered when the ligand has no symmetry as RMSD/ub matches each atom in a conformation with itself in the other conformation ignoring any symmetry. We have used RMSD/ub as the ligand molecules have no specific symmetry. Another criteria of choosing best ligand: protein pose is identifying the types and number of bonding between them. The metabolite which makes maximum number of H-bonds, hydrophobic bonds with the receptor protein mostly show better capability to form ligand: protein complex formation by intermolecular interaction between inhibitor and receptor. Different output poses were analysed in Discovery Studio visualizer 2020 version 20.1.0.19295 (Dassault Systèmes BIOVIA) for the formation of non-bonded hydrogen bonds. The best pose structure was analysed also by their binding affinity, inhibition constants and other supporting interactions.

| *Drug likeness properties* | **berberine (a)** | **choline (b)** | **β-sitosterol (c)** | **tetrahydropalmatine (d)** | **octacosanol (e)** |
|---|---|---|---|---|---|
| | ***Physicochemical Properties*** | | | | |
| ***Molecular weight (gm/mol)*** | 336.361 | 104.1708 | 414.718 | 355.434 | 410.7595 |
| ***Num.H-bond acceptors*** | 4 | 1 | 1 | 5 | 1 |
| ***Num. H-bond donors*** | 0 | 1 | 1 | 0 | 1 |
| ***No of rotatable bonds*** | 2 | 2 | 6 | 4 | 26 |
| ***Molar Refractivity*** | 94.87 | 30 | 133.23 | 103.99 | 137.87 |
| ***Topological polar surface area TPSA (Å$^2$)*** | 40.80 Å$^2$ | 2 0.23 Å$^2$ | 20.23 Å$^2$ | 40.16 Å$^2$ | 20.23 Å$^2$ |
| | ***Lipophilicity*** | | | | |
| ***Log $P_{o/w}$ (iLOGP)*** | -0.00 | -2.41 | 5.05 | 3.69 | 7.20 |
| ***Log $P_{o/w}$ (XLOGP3)*** | 3.62 | -0.40 | 9.34 | 3.24 | 13.61 |
| ***Log $P_{o/w}$ (WLOGP)*** | 2.19 | -0.32 | 8.02 | 2.52 | 10.14 |
| ***Log $P_{o/w}$ (MLOGP)*** | 2.53 | -3.46 | 6.73 | 2.20 | 7.07 |
| ***Log $P_{o/w}$ (SILICOS-IT)*** | 7.24 | -0.57 | 7.04 | 3.75 | 10.96 |
| ***Concensus Log Po/w*** | 9.80 | -1.38 | 7.24 | 3.08 | 9.80 |
| | ***Druglikeness*** | | | | |
| ***Lipinski*** | yes | yes | yes | yes | yes<br>(1 violation) |
| ***Veber*** | yes | yes | yes | yes | yes<br>(1 violation) |
| ***Ghose*** | yes | Partly yes (2 violation) | No | yes | No |
| ***Egan*** | yes | yes | yes<br>(1 violation) | yes | yes<br>(1 violation) |
| ***Muegge*** | yes | yes (1 violation) | Partly yes<br>(2 violation) | yes | No |
| ***Bioavailability score*** | 0.55 | 0.55 | 0.55 | 0.55 | 0.55 |
| ***Synthetic accessibility (SA)*** | 3.14 | 1 | 6.30 | 3.59 | 3.72 |
| | ***Pharmacokinetics*** | | | | |
| ***GI absorption*** | High | Yes | Low | High | Low |
| ***BBB permeant*** | Yes | No | No | Yes | No |
| ***P-gp substrate*** | Yes | No | No | Yes | Yes |
| ***Log $K_p$ (skin permeation)*** | -5.78 cm/s | -7.22 cm/s | -2.20 cm/s | -6.17 cm/s | 0.86 cm/s |
| | ***Water Solubility*** | | | | |
| ***Log S (SILICOS-IT)*** | -5.92 | -1.26 | -6.19 | -5.87 | -10.53 |
| **Solubility (mg/mL)** | 4.00e-04 | 5.74e+00 | 2.69e-04 | 4.81e-04 | 1.20e-08 |
| **Toxicity** | Rat LD50: 2.7834 mol/kg | Oral rat $LD_{50}$: 3400 mg/kg | Not Available | Not Available | Not Available |

**Table 1(a-e):** Molecular configuration and drug likeliness properties of proposed ligand drug molecules for COVID-19 by SWISS ADME data.

## 2.5. MD Simulations

At atomistic level, MD simulation results helped us to investigate the structural dynamics of receptor CoV-2 protease (3CL$^{pro}$) upon binding with small ligand (proposed drug) molecule. LINUX based platform ‘‘GROMACS 5.1 Package’’ (Berendsen et al., 1995) was used for determination of thermodynamics stability of the proposed ligand: protein complex. MD simulations were performed using latest CHARMM36 all atom force field (Soteras et al 2016). Before simulation, separate topologies were prepared for receptor and ligand by different external tools (Gajula et al 2016). For our ligand, before creating the topology first we have optimized the ligand structure by Gaussian 9.0 by Density Functional Theory (DFT) with the basis set 6.31G (d,p) (Becke et al 1997, Frisch et al 2004). With the optimized structure we have generated the ligand topology with the CHARMM General Force Field (CGenFF) program (Vanommeslaeghe et al 2012). In CHARMM all-atom force field, all H atoms of ligand are explicitly represented. For topology we have used the cgenff_charmm2gmx.py (http://mackerell.umaryland.edu/charmm_ff.shtml#gromacs). To perform simulations in aqueous solution, we have used a well known water model: TIP3P. During solvation process the bare protein and protein: ligand complex were solvated in the cubic box having specific boundary conditions (with 10 Å buffer distance) with volume as 893,000 Å$^3$. As per procedure 4$Na^+$ ions were added for electrically neutralizing the probe system. Before simulation we have done the energy minimization on the system to sort out any bad starting structures and also to minimize solute structure in vacuum before introducing solvent molecules. The Protein ligand complex system was equilibrated under suitable simulation parameters consistent for our selected CHARMM General Force Field for building the energy minimized solvated system. The steepest descent algorithm has been used for energy minimization of the system with varying time (ps) for 500,000 iteration steps. For reducing the steric clashes, the applied algorithm has a cut-off up to 1000 kJmol$^{-1}$ (Kumar et al 2020). The MD simulation has started with the minimization of the system. The minimization has been achieved at two phases each having 500,000 steps. In first part, equilibration was obtained having each step of 2 fs with a boundary condition of constant number of particles (N), volume (V), and temperature (T). We have used 10000-ps *NVT* equilibration. In second phase, the equilibration was achieved under the pressure of 1 atmosphere 298K. Here the boundary condition maintained as constant NPT (particle numbers, pressure, temperature). Such condition is known as isothermal-isobaric ensemble. Energy minimization data and output of MD simulation data helped us to visualize several thermodynamic parameters like potential energy, kinetic energy, total energy, temperature, density progression, radius of gyration, RMSD, RMSF, SASA, etc. of the bare host protein and ligand: receptor complex. For equilibration step computation, for covalent bond constraints LINCS algorithm was applied. To quantify strength of interaction between ligand and protein it is useful to compute the nonbonded interaction energy between two species rather than calculating free energy of the system. For the nonbonded energy calculation we have used short range Lennard-Jones and Coulomb interaction energies. For these calculation we have maintained a 1.4 nm radius cut-off. For long range electrostatics calculation Particle Mesh Ewald (PME) method was used having 1.6Å of Fourier grid spacing. Berendsen temperature coupling method (V-rescale ) has been used for maintaining inside box temperature. NPT equilibration has been achieved by Parrinello-Rahman pressure coupling method. The time parameter has been fixed to 10000 ps with each step of 0.002 fs for the final step of simulation. After final step of each simulation, trajectories were obtained. Obtained trajectories and results were analysed using the graphical tool Origin pro. Least-square fitting method was used to evaluate RMSD for protein backbone. Similarly, RMSF was obtained for protein $C_\alpha$ backbone. The compactness factor of protein was calculated by radius of gyration ($R_g$). According to rule if the protein or complex structure is stable then the radius of gyration should maintain a stable value. To compute total solvent accessible surface area (SASA) the tool sasa was used. SASA measures the area of receptor exposure to the solvents during the simulation process. Number of hydrogen bonds and distribution of intermolecular hydrogen bond lengths were calculated with maintaining 3.5Å distance cut-off condition throughout the simulation.

## 2.6. Computational Details

Dell Gen9 server with 8 Core I7 processors and 16 GB of RAM with GPU NVIDIA MX130 was used for the MD simulations and corresponding energy calculations.

# 3. Results and Discussion

## 3.1. Virtual screening and Molecular Docking analysis

Each expected ligand drug molecule was docked to CoV- 2 3CL main protease ($3CL^{PRO}$). We have to keep in mind that docking algorithm is nondeterministic in nature. The minimum of scoring function cannot be identified by docking algorithm for a fixed receptor: ligand complex conformation. Table SD1 shows the structure of possible ligands found in the active site pockets of $3CL^{PRO}$ proteins I and II. Molecular docking is used to find out interaction of tested inhibitors: berberine, β-sitosterol, choline, tetrahydropalmatine and octacosanol with both $3CL^{PRO}$ proteins I and II. The organic compound berberine ($C_{20}H_{18}NO_4$) can be extracted from Hydrastis canadensis L., Berberidaceae. It is also known as protoberberine alkaloids. Berberine is usually found in the roots, rhizomes, stems, and bark. One of the main source of berberine is *Tinospora cordifolia.* It can be found in other plants also. Berberine is used as a natural dye with a color index of 75160. Due to its strong yellow fluorescence, it is useful in histology for staining heparin in mast cells. Berberine is relatively toxic parenterally, but still it is used as oral drug for the treatment of various parasitic and fungal infections. It is also used a drug for antidiarrheal diseases (Dewick et al 2009). β-sitosterol ($C_{29}H_{50}O$) has a structural similarity with cholesterol. It is used to reduce cholesterol levels in the body. It is used as a constituents in many drugs for reduction of swelling. Being a steroid, β-sitosterol is a precursor of anabolic steroid boldenone (Wang et al 2008). Octacosanol ($C_{28}H_{58}O$) is basically an alcohol "straight-chain aliphatic 28-carbon primary fatty alcohol" which is used health industry as a nutritional supplement. It has been studied as a potential therapeutic agent for the treatment of Parkinson's disease (Rudkowska et al 2008). Octacosanol is reported to possess cholesterol-lowering effects, antiaggregatory properties and ergogenic properties. Tetrahydropalmatine ($C_{21}H_{25}NO_4$) has a role as an adrenergic agent, a non-narcotic analgesic and a dopaminergic antagonist. Tetrahydropalmatine is under investigation in clinical trial for the Treatment of Schizophrenia (Taylor et al 2003). Choline ($C_5H_{14}NO$) is a nutrient that supports various bodily functions, including cellular growth and metabolism. It is also used for DNA synthesis and nervous system maintenance (Corbin et al 2012).

Candidate drugs that confirm to the Ro5 and other drug likeness rules etc. used to have lower attrition rates during clinical trials and hence have shown their strong candidature as potential drugs reaching the market (Table 1a-e). Berberine, choline and tetrahydropalmatine were satisfying the most prescribed virtual screening properties. Auto Dock Vina, the docking software uses free-energy scoring function based on a linear regression analysis to identify the best binding mode prediction for protein: ligand complexation. It identifies the best possible pose structure from a set of diverse protein-ligand complexes with a highest negative value of binding energy and lowest value of inhibition constant. Docking simulation results have helped us to find out the best possible interacted protein: ligand poses of different proposed inhibitor drugs. For the first tested inhibitor berberine with $3CL^{pro}$ protein I (6LU7), berberine: 6LU7 complex revealed 9 different poses based on their binding affinity, drieding energy, dipole moment, inhibition constant etc. According to molecular docking simulation protocol the pose with highest negative values of binding energy, lowest value of complex driending energy usually indicates maximum binding affinity for the ligand: receptor complex formation (Table 2).

| Pose | Binding affinity (G) (kcal/mol) | Hydrogen bonded interaction number | Drieding energy (ligand) | Drieding energy (protein) | Dipolemoment of ligand(debye) |
|---|---|---|---|---|---|
| 1 | -7.3 | 0 | 414.419 | 6,044.79 | 1.676 |
| 2 | -7.3 | 2 | 415.725 | 6,045.91 | 1.677 |
| 3 | -7.1 | 0 | 412.076 | 6,042.69 | 1.675 |
| 4 | -7.0 | 0 | 414.361 | 6,044.95 | 1.68 |
| 5 | -7.0 | 0 | 415.549 | 6,045.57 | 1.672 |
| 6 | -6.9 | 0 | 414.623 | 6,044.94 | 1.676 |
| 7 | -6.9 | 0 | 414.409 | 6,044.77 | 1.68 |
| 8 | -6.9 | 0 | 415.07 | 6,046.38 | 1.68 |
| 9 | -6.8 | 0 | 415.635 | 6,046.05 | 1.675 |

**Table 2.** Binding mode of each ligand:protein (Berberine: 6LU7) complex using molecular docking for Protein (I).

For pose 2 we obtained the better interacted position for ligand: protein complex with the binding affinity of -7.3 kcal/mol which is lower than highly tested COVID 19 drug chloroquine (-6.29 kcal/mol). To verify that this the best docked site we have also computed the Dreiding energy of different poses which consider the Dreiding force field to calculate the energy of a specific complex structure by summing energy components like bond lengths, bond angles, dihedral angles (Mayo et al 1990). The Dreiding energy values for individual protein and ligand are different but when they form complex the Dreiding energy becomes minimum for most favourable structure. We obtained the lowest value of energy (415.725) for the pose 2 of the complex structure (Table 3). To validate the better interaction between berberine and 6LU7, we have computed the inhibition constant ($k_i$) which is an indication of how potent berberine as inhibitor towards 6LU7. It is the concentration to produce half maximum inhibition. To evaluate the inhibition constant we have used relation as

$$k_i = e^{\left(\Delta G / RT\right)} \quad \ldots\ldots\ldots\ldots(1)$$

Where G is the binding affinity, R the universal constant and T (298 K) the room temperature.

For the best pose (2) structure of berberine: 6LU7 complex, the $k_i$ value obtained at room temperature (298 K) as 4.4 x$10^{-6}$ M which proves the higher affinity of berberine towards receptor 6LU7 (Table 3). The computed $k_i$ value is far lower than the toxicity dose range of berberine as proposed drug (Table 1) which validate the strong candidature of berberine as target drug to attach 6LU7. The berberine: 6LU7 interaction energy further verified by the dipole moment values of ligand and target (Table 3). The strong interaction for pose 2 was further verified by the number of weak nonbonded hydrogen bonded interactions and hydrophobic interactions present between protein and ligand in the optimized protein: ligand complex structure. Hydrogen bonded interaction and hydrophobic interaction due to their weak nature have significant roles in stabilizing the energetically-favored ligands in a suitable pocket of the environment of protein structures. It is a proven fact that weak interactions "hydrogen bonding and hydrophobic interactions" always stabilize the ligands at the target protein site by altering the binding affinity (Patil et al 2010). We observed the presence of intermolecular hydrogen bonds and hydrophobic interaction between protein (Residues: THR25, SER46, HIS163) and berberine in its pose 2 which means better interaction between donor and acceptor moiety. We have repeated the molecular docking simulation of berberine with other $3L^{pro}$ protein II also and obtained the best ligand: protein complex structure with binding affinity of -7.7 kcal/mol and inhibition constant of 2.23 $x10^{-6}$M (Table SD3). For best poses of berberine: protein complexs, the donor–acceptor surface and different possible interactions are shown in Figure (4a, b) in 3D and 2D view.

| Ligand | Best Binding affinity (kcal/mole ) | Hydrogen bonded interaction (protein donor: ligand acceptor, distance in Å) | Hydrophobic interaction (protein donor: ligand acceptor, distance in Å) | Dipole moment of ligand (debye) | Drieding energy between protein and ligand | Inhibition constant (M) |
|---|---|---|---|---|---|---|
| berberine | -7.3 | (A:THR25:HG1 - :UNK0:O, 2.11198)<br>(A:SER46:HG - :UNK0:O, 2.88545) | (:UNK0 - A:HIS163, 3.61032) | 1.67 | 415.725 | 4.4 $x10^{-6}$ |
| β-sitosterol | -7.1 | (:UNK0:O - A:ARG188:O, 3.12606) | (:MET165 -:UNK0, 4.54304) | 1.761 | 1280.56 | 6.16 $x10^{-6}$ |
| choline | -3.4 | (:UNK0:O - A:TYR54:OH, 2.8027)<br>(:UNK0:C - A:MET49:O, 3.71379)<br>(:UNK0:C- A:GLN189:OE1, 3.61393) | (:UNK0:C - A:HIS41, 3.8048) | 5.374 | 104.198 | 3.2 x $10^{-3}$ |
| tetrahydropalmatine | -6.4 | | (:UNK0:C - A:MET165, 4.73002)<br>(:UNK0:C - A:LEU167, 4.89871)<br>(:UNK0:C - A:PRO168, 5.01258)<br>(:UNK0 - A:CYS145, 5.27402)<br>(:UNK0 - A:MET165, 4.58548) | 2.688 | 373.677 | 2.01 x $10^{-5}$ |
| octacosanol | -6.6 | (:UNK0:O - A:LEU141:O, 3.37559) | (:UNK0 - A:ALA191, 4.8791) | 1.257 | 414.094 | 1.43 x $10^{-5}$ |
| | | (:UNK0:C - A:HIS163:NE2, 3.62548) | | | | |

**Table 3:** Interaction detail for different ligands: berberine, β-sitosterol, choline, tetrahydropalmatine and octacosanol with receptor protein I (6LU7).

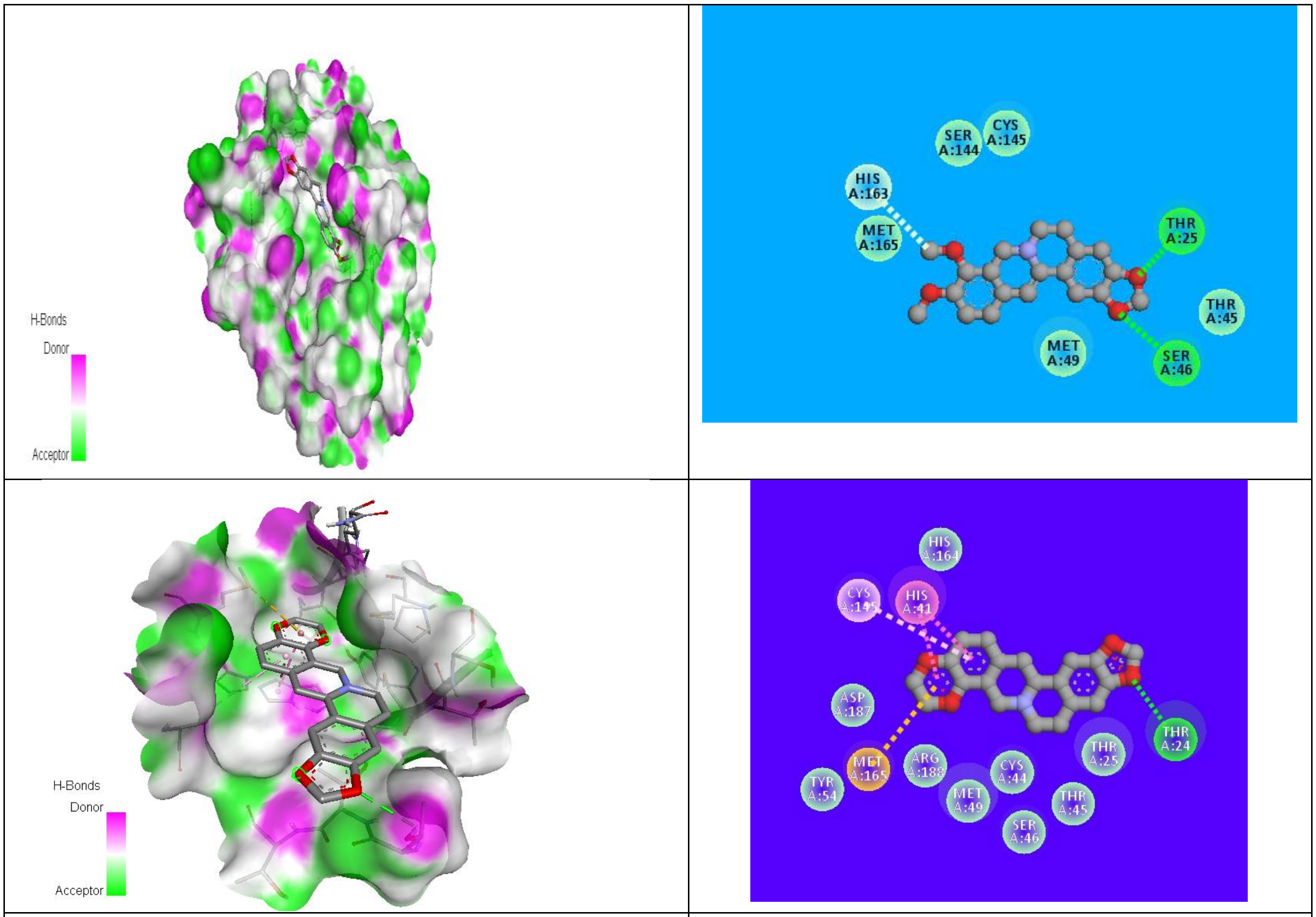

| | |
|---|---|
| a. 3D view of ligand: protein interaction surface due to hydrogen bond.<br><br>Top row = berberine: 6LU7<br>Lower row = berberine: 6W63 | b. 2D view of ligand: protein interaction<br>Green: conventional hydrogen bond<br>Light pink: π-alkyl bond<br>Deep pink: π-π bond<br>Sky blue: π-donor hydrogen bond<br>Yellow: π-Sulphur bond |

**Figure 4a,b:** Donor: acceptor surface and possible types of interactions in best pose structure obtained from molecular docking for berberine: 6LU7 and berberine: 6W63

We have repeated same molecular docking approach for other ligand structures: β-sitosterol, octacosanol, Tetrahydropalmatine, Choline with protein I, protein II and have identified their best possible ligand: protein interaction pose position in terms of their best binding affinity value, Dreiding energy, dipole moment, inhibition constants, number of hydrogen bond, hydrophobic bond etc. and mentioned them all in Table 3,SD3 and Figure (5a,b) and Figure SD2 in 3D and 2D view. From the molecular docking results for every suggested drug molecule, with their best possible ligand: target complex structures i.e., for best pose, the donor–acceptor surface with their possible hydrogen bonding and hydrophobic interactions are shown in Figure (SD2a,b) in 3D and 2D view. Our screening identified that out of five possible bioactive ligand structures, berberine shows the best potentiality to inhibit with the CoV-2 enzymes ($3L^{pro}$): I and II by its best docking affinity compared to the other ligands against $3L^{pro}$. Berberine: $3L^{pro}$ complex has shown less binding energy, minimum inhibition constant value as compared to other proposed drug molecules with its good binding mode of interactions. After berberine the binding affinity followed by β-sitosterol, octacosanol, Tetrahydropalmatine, Choline. Simulation results also revealed that all five proposed drug molecules showed good stability as a complex with the targeted $3L^{pro}$ protease. These inhibitors also satisfy the required drug likeness properties according to Ro5, Veber etc rules including their molar refractivities, pharmakokinetics, polar surface areas and logP values. All of the reported potential natural phytoconstituent drug compounds are commercially available in drug industry and so can be easily available for further in vivo/in vitro validations. We are hopeful that the information generated from this present study will definitely be utilized for the development of phytochemical based therapeutics against COVID-19. Since berberine shows the better possibility of inhibition towards $3CL^{pro}$ protease we have further studied the applicability of berberine as potential drug molecule by applying rigorous MD simulation approach.

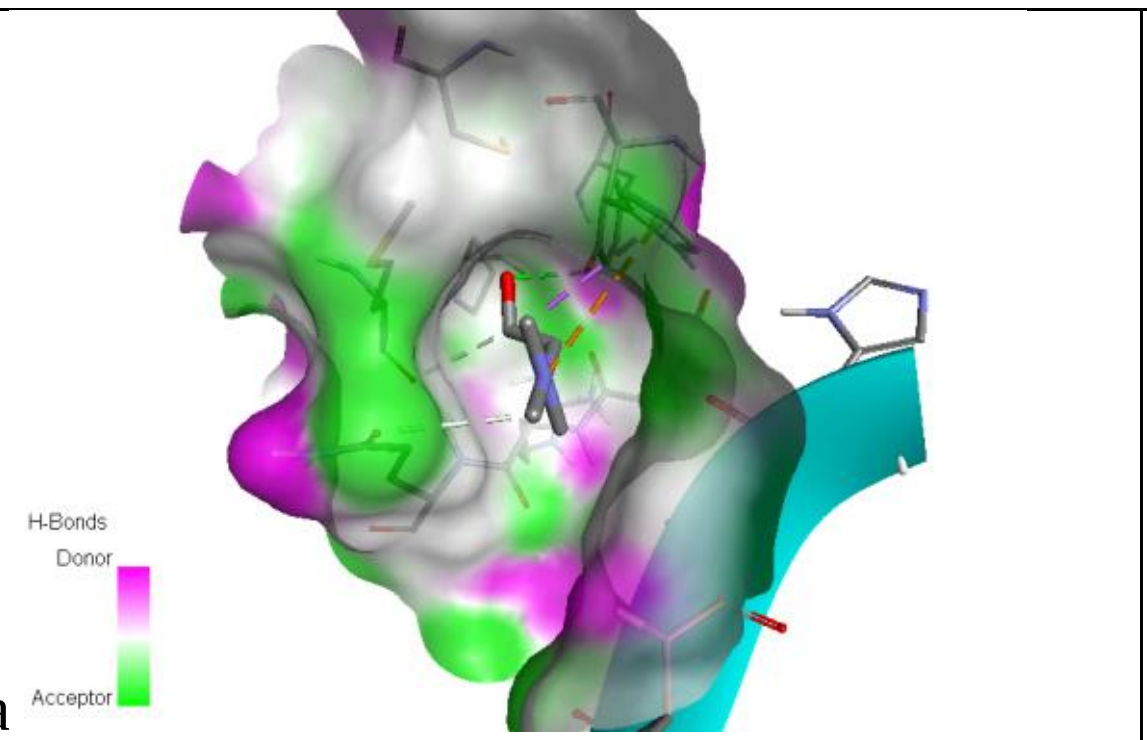

a

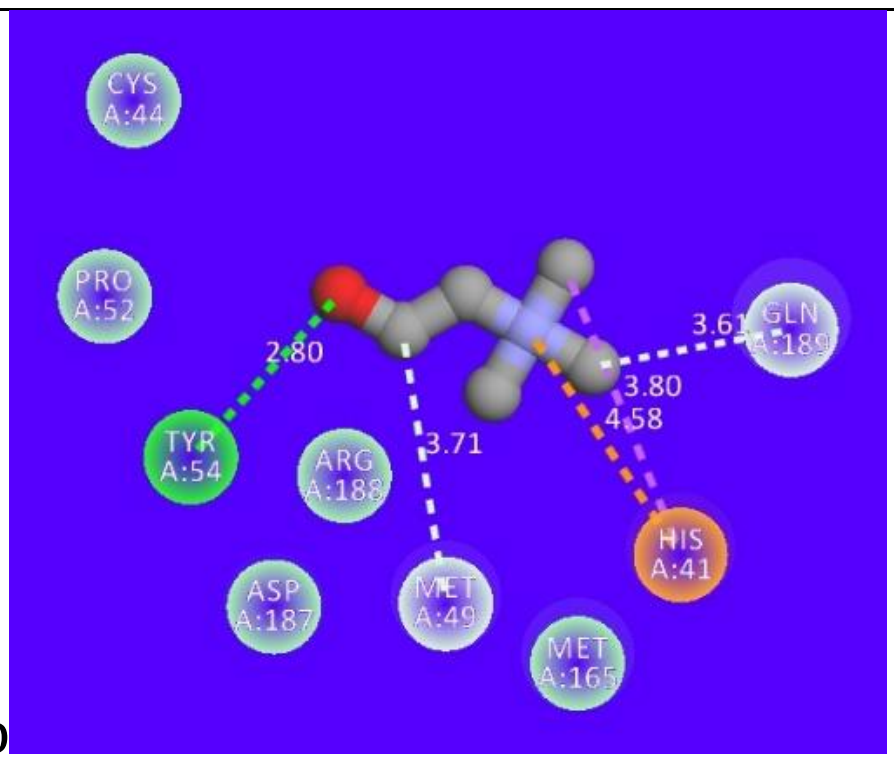

b

| 3D: Choline: 6LU7 | 2D: Choline: 6LU7 |
|---|---|

<table>
<tr><td>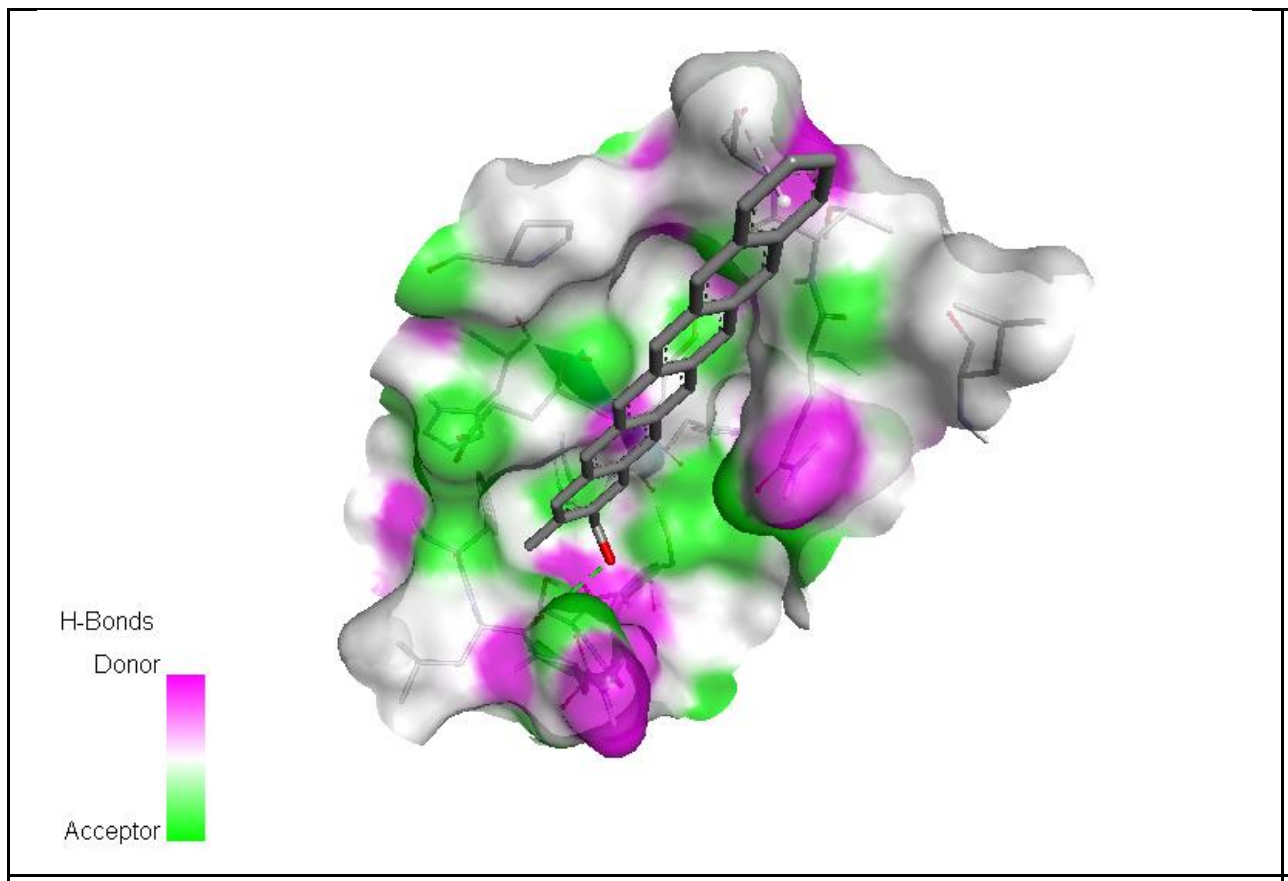<br></td><td>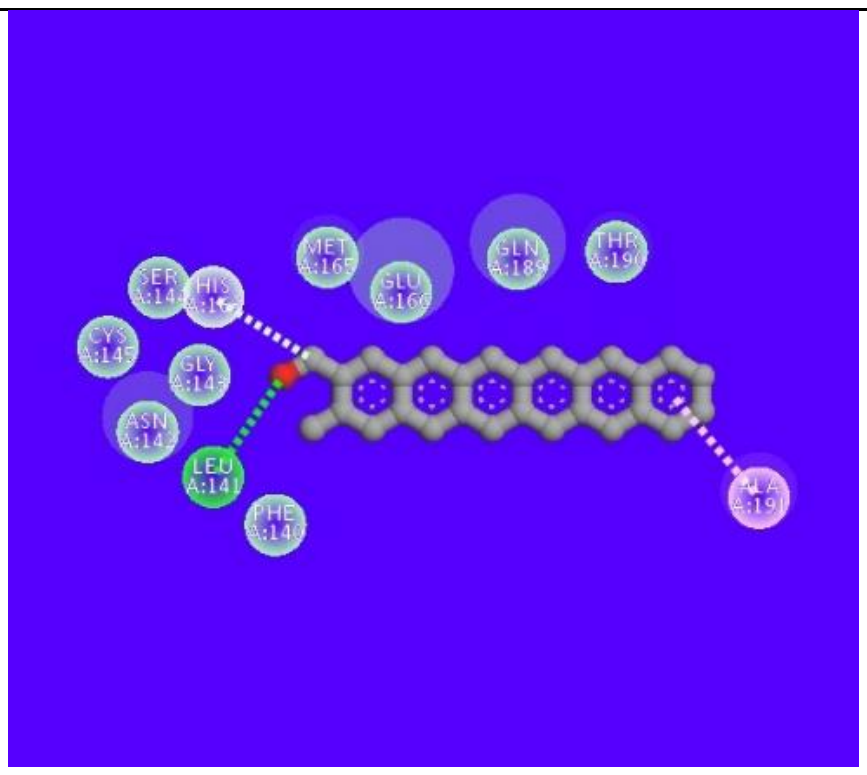</td></tr>
<tr><td>3D: octacosanol: 6LU7</td><td>2D: octacosanol: 6LU7</td></tr>
<tr><td>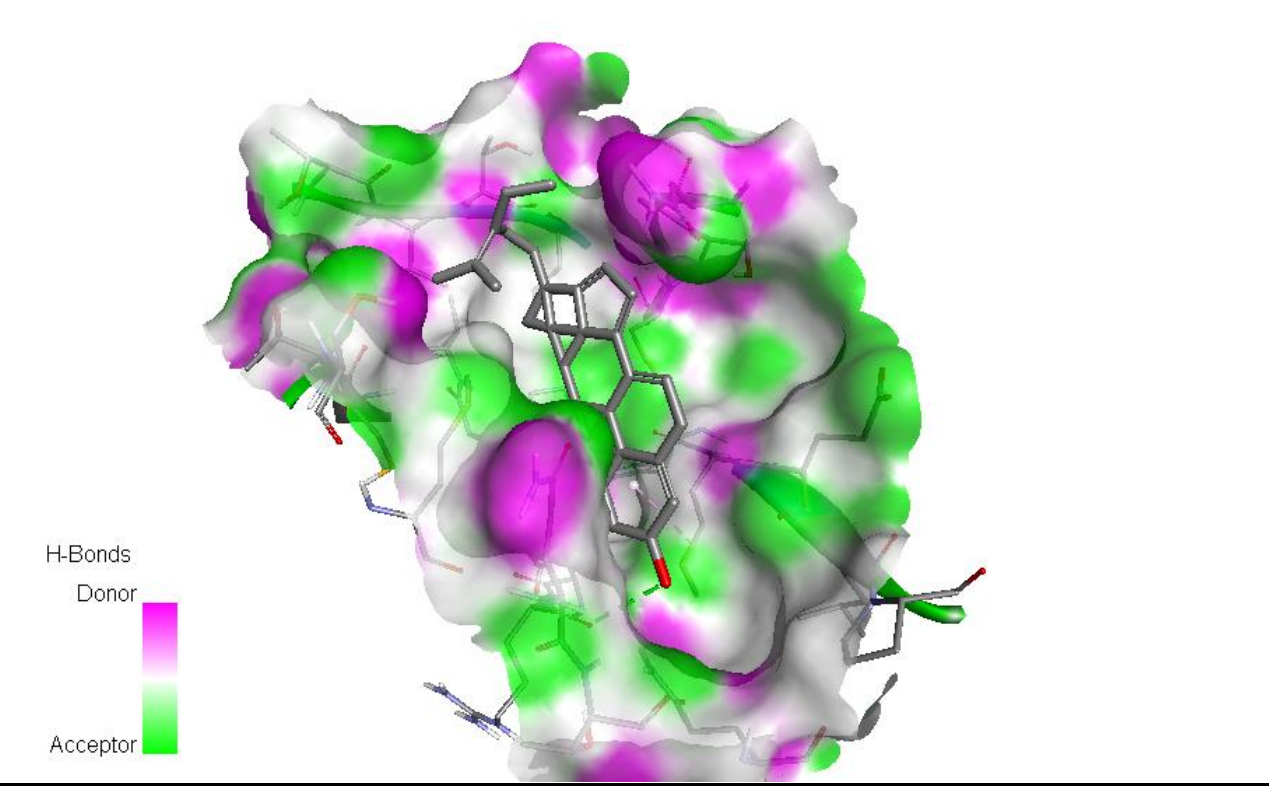<br></td><td>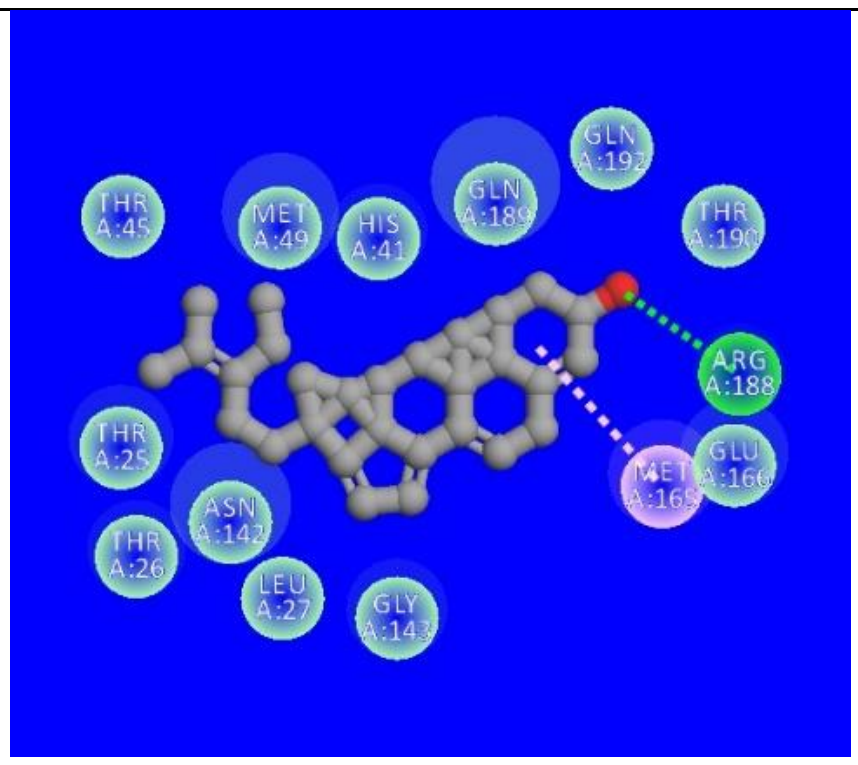</td></tr>
<tr><td>3D: β-sitosterol: 6LU7</td><td>2D: β-sitosterol: 6LU7</td></tr>
<tr><td>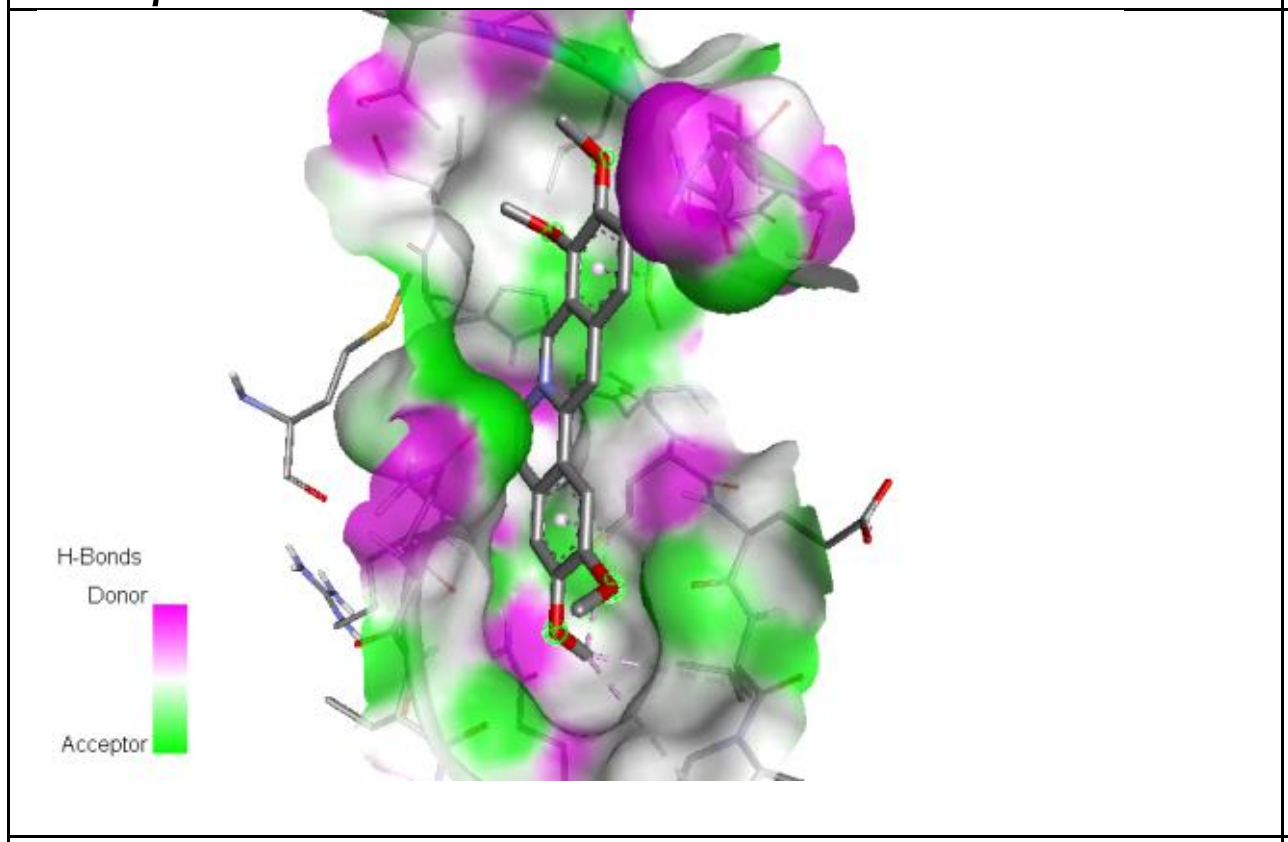<br></td><td>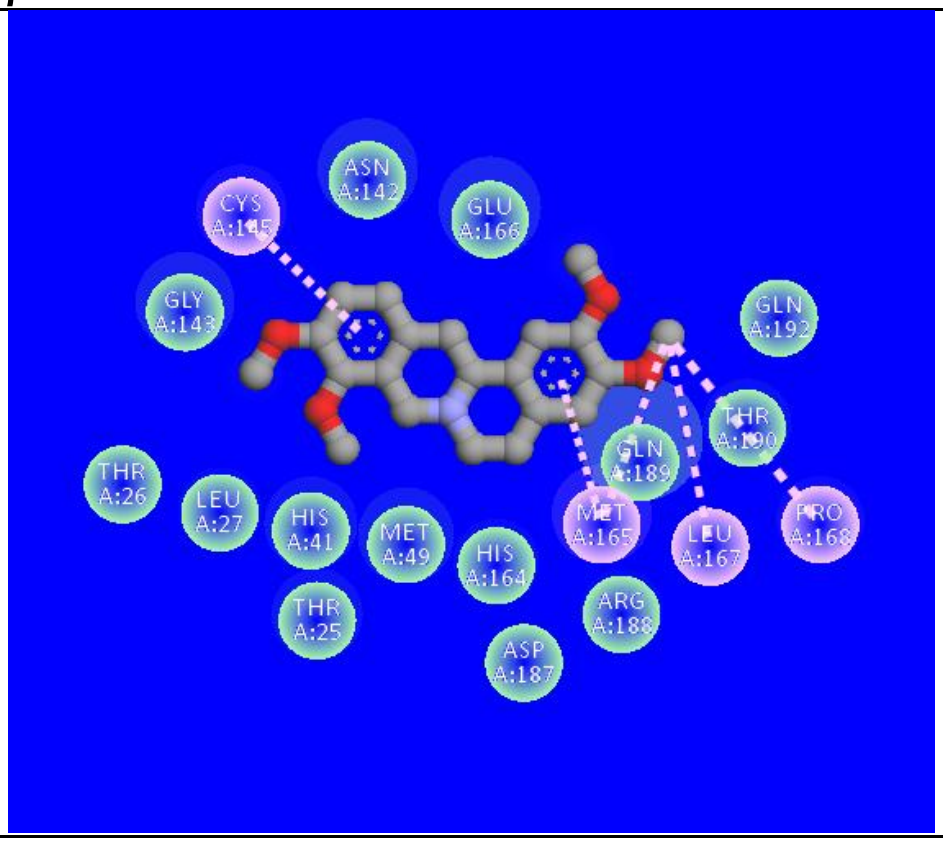</td></tr>
<tr><td>3D: tetrahydropalmatine: 6LU7</td><td>2D: tetrahydropalmatine: 6LU7</td></tr>
<tr><td>a. 3D view of ligand: protein interaction surface due to hydrogen bond.</td><td>b. 2D view of ligand: protein interaction<br>Green: conventional hydrogen bond<br>Light pink: π-alkyl bond<br>Deep pink: π-π bond<br>Sky blue: π-donor hydrogen bond<br>Yellow: π-Sulphur bond</td></tr>
</table>

**Figure 5.** Donor: acceptor surface and possible types of interactions in best pose structures obtained from molecular docking for different ligands with COVID-19 protease enzyme 3CL$^{pro}$ (6LU7).

### 3.2 MD Simulation analysis

For obtaining dynamic data at atomic spatial resolution, MD simulation is a verified *in silico* method which can simulate in picoseconds/nanoseconds or further finer temporal steadfastness (Benson & Daggett, 2012; Gajula et al., 2016). The complex (berberine: 6LU7) formed by main protein protease (6LU7) docked with phytochemical compound (berberine) has been studied under this simulation for 10000 ps to 100 ps period to analyze the stability of the studied structure. For MD simulation to work every structure (receptor and complex) have to be optimized by energy minimization process. For energetically minimized structure the potential energy should be minimum and negative with a maximum force value. We have obtained energetically minimized structures for both probe (protein and complex) systems (Figure. 6). For both the cases we have obtained steady convergence of potential energy. The analyzation and comparison of the potential energy ($E_{pot}$) of the stabilized structure of 6LU7 protein in its bare state and in berberine: 6LU7 complex form have been done rigorously. 6LU7 has an average energy of $-1.27\text{x}10^6$ ±56.7 (kJ $mol^{-1}$ ) for $E_{pot}$, while the complex has an averaged $E_{pot}$ of $-0.2 \text{ x } 10^6 \pm 38.77$ (kJ $mol^{-1}$ ). With their lowest $E_{pot}$ values both systems at an energy minimum level and ready real Molecular Dynamics. To check further, optimized structures of $3CL^{pro}$ protease and complex structures have been equilibrated by NVT and NPT ensembles to stabilized temperature, pressure, density, volume etc within a time scale trajectory of between 100 ps – 10000 ps. Analysing the temperature progression data (Figure SD4) with an running average value of 10 ps, it was observed that after starting the simulation process, the temperature of the system quickly reached the stable room temperature value (298 K) with maintaining the temperature stability throughout the whole equilibration process. This is the reason for bare protein system, a shorter equilibration period (on the order of 50 ps/100 ps) have been applied. The applicability of shorter equilibration period have been also verified by pressure and density data (Figure SD5) with same running average value of 10 ps. Also we have observed a very stable values of density and pressure over the period of time trajectory. From the stability of temperature, pressure and density values it was justified to conclude that the system is well-equilibrated with respect to temperature, pressure and density and are ready for MD simulation. We have studied various thermodynamic parameters of host protein and complex system by MD simulation data for various time trajectory. The important thermodynamic parameters we have studied are the RMSD, RMSF, inter-molecular H- bonds, $R_g$, SASA, potential energy, kinetic energy, total energy and various nonbonded interaction energies to understand the stability and possible conformational changes of the bare state of $CL^{pro}$ receptor protein and in the berberine: 6LU7 complex at each frame of the time resolved simulation trajectory. All MD simulated output of bare protein and protein: ligand complex for the time resolved trajectories have been presented in Table 4.

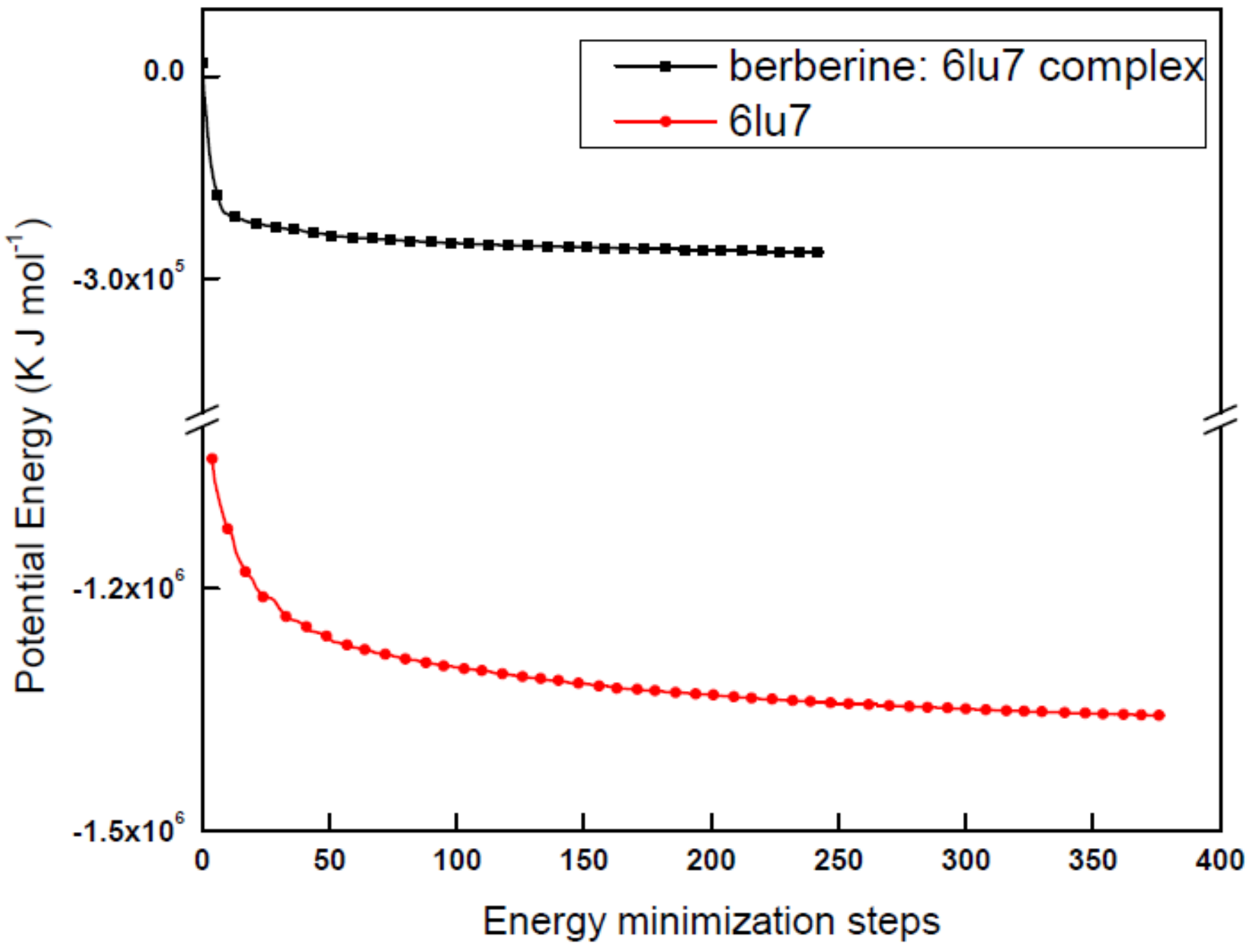

**Figure 6:** Optimized geometry of receptor 6LU7 and berberine: 6LU7 complex.

| Serial no | Parameter | Bare 3CL$^{pro}$ protease (6LU7) | | 6LU7 + berberine complex | |
|---|---|---|---|---|---|
| | | Mean | Range | Mean | Range |
| 1. | SR Columbic Interaction energy (kJ mol$^{-1}$ ) | NA | NA | -51.7714± 3.8 | 20 - -80 |
| 2. | SR LJ Interaction energy (kJ mol$^{-1}$ ) | NA | NA | -130.261± 1.1 | -80 - -160 |
| 3. | Average Interaction Energy energy (kJ mol$^{-1}$ ) | NA | NA | -186.95± 5.1 | |
| 4. | RMSD (nm) | 0.12 | 0.08 – 0.16 | 0.16 | 0.08 – 0.18 |
| 5. | Inter H-Bonds | NA | NA | 0.026 | 0 - 2 |
| | Intra H-Bonds | 630 | 614 - 653 | 213 | 196 -220 |
| 6. | Radius of gyration | 2.25± 0.01 | 2.25 – 2.26 | 2.25± 0.01 | 2.25 – 2.26 |
| 7. | SASA (nm$^2$) | 22 | 19 - 26 | 148 | 146 - 156 |
| 9. | Potential Energy (kJ mol$^{-1}$ ) | -1.27x10$^6$ ±56.7 | -7.3 x10$^5$ - -1.3 x10$^6$ | -0.2 x 10$^6$ ± 38.77 | -2.4x10$^4$ - -0.2x10$^6$ |
| 10. | Total Energy (kJ mol$^{-1}$ ) | -1.17 x10$^6$±0.01 x 10$^6$ | -1.3 x10$^6$ - -1.0 x10$^6$ | -1.8 x10 $^5$ ± 0.01 x 10$^5$ | -1.8 x10 $^5$ - -1.7 x10$^5$ |

**Table 4**. MD simulation output of time resolved trajectory of 6LU7 in its bare state without any ligand and in the complex state with ligand berberine.

Throughout the whole time resolved trajectory the total energy of both bare protein (6LU7) and berberine: 6LU7 complex remain stable having average value of -1.8 x10 $^{5}$ kJ mol$^{-1}$ with a minor fluctuations between-1.8 x10 $^{5}$ kJ mol$^{-1}$ and-1.7 x10$^{5}$ kJ mol$^{-1}$ which validate the good stability of both bare protein and its complex structures for the whole time trajectory (Figure SD 6,7). The variation of total energy of the complex due to its kinetic energy and potential energy components is shown in Figure 7.

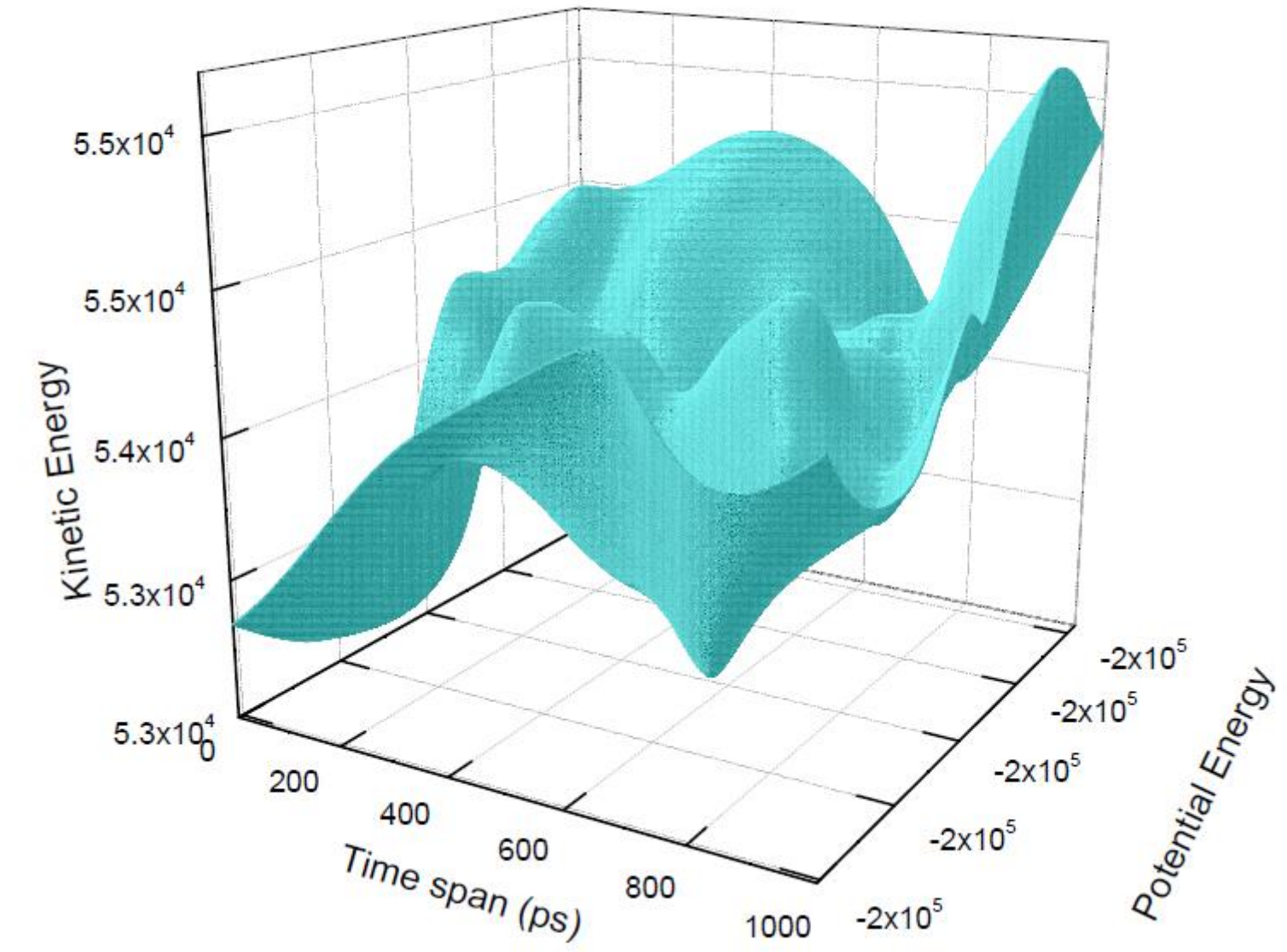

**Figure 7:** Total energy of berberine: 6LU7 Complex structure in time scale 1000 ps.

Protein backbone RMSD (Figure. 8) of bare protein was observed to be fluctuating between 0.08 and 0.16 nm with an average value of 0.12 nm. In presence of compound berberine after complex formation berberine: 6LU7, RMSD showed a stable and constant range between 0.08 nm to 0.18 nm with an average value of 0.16 nm. The obtained average RMSD for complex is comparatively slightly higher than the RMSD of bare protein backbone. The complex RMSD data and bare background protein have been compared with the crystal structures for both the configuration. With full time scale range the obtained RMSD variation levels up to ~0.1 nm for both bare protein and protein: ligand complex indicating the stability of both the structures. Subtle differences between the plots of gaseous structure and crystal structure indicated minor structural difference at t = 0 ps. This minor difference in structures is very much expected since for energy-minimized structures position restraints can never be 100% perfect. Comparing the differences of protein backbone RMSDs for bare $CL^{pro}$ protein structure with berberine: protein complex structure suggest that $CL^{pro}$ has shown no significant change in presence of berberine during time resolved simulation. Whereas the RMSD of complex (berberine: 6LU7) has indicated several fluctuations at different time intervals on 10000 ps time scale. These fluctuations may arise due to the changed conformation in binding region of $CL^{pro}$ (Figure SD8). In the full time scale intervals, the first stable structure was observed between 1645 ps to 2860 ps, second one is observed between 3574 ps to 6372 ps and the third one is between 6523 ps to 8753 ps. Throughout the whole time trajectory, in the stable region the RMSD remain constant at 0.5 nm whereas for large fluctuation regions the observed RMSD was more than 0.6 nm. The fluctuations were observed between 0 ps to 1644 ps, 2861 ps to 3573 ps and 6373 ps to 1000 ps.

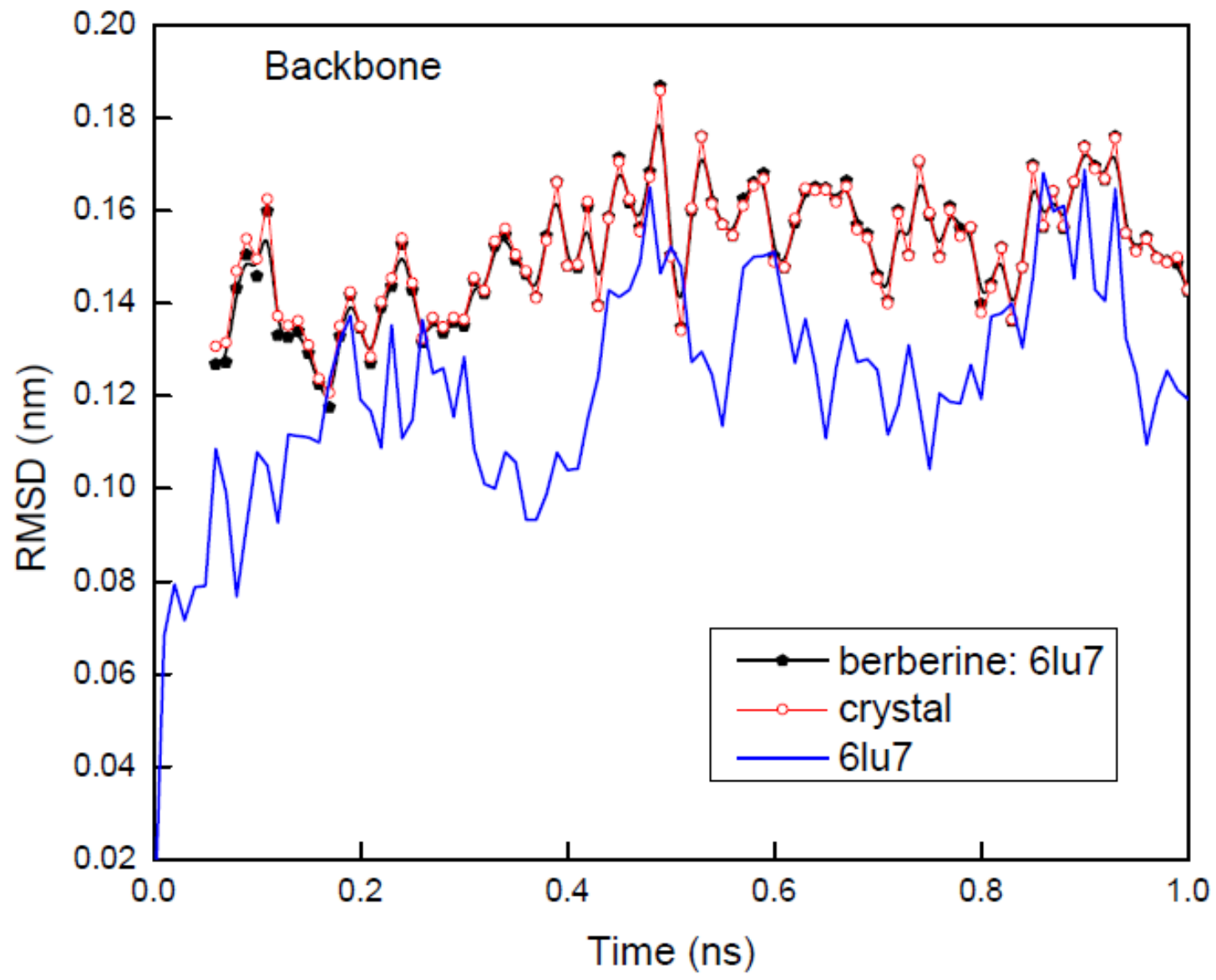

**Figure 8:** Root mean square Deviation (RMSD) backbone graphs of 6LU7 in its bare state and in complex with berberine. Complex structure is compared with its crystal form also.

The above observation also confirmed by the RMSF plots of $C_\alpha$ for both the bare protein and its complex where local changes were observed with variation of residues numbers (Figure 9). For bare protein 6LU7, RMSF value for each residue was simulated and averaged for the entire time trajectory and identification has been done for more flexible regions responsible for the conformational changes. In simulated RMSF graph of bare 6LU7, major fluctuations in peaks have been observed with the residues positions between 50-80 and the tail region of the protein with the residue number around 280. The fluctuations in RMSF values were observed between 0.13 nm and 0.23 nm. For berberine: 6LU7 complex, the major fluctuations in peaks were observed initially around residue number at 50 and in between140–200 values, with fluctuations around 0.23 nm. The RMSF data suggest that all active and flexible residues of 6LU7 get stabilized due to binding of berberine for the complex formation, which hints the possibility of inhibitory activity of the ligand towards receptor protein without affecting the protein structure. From the detailed RMSD and RMSF analysis we can conclude that protein backbone of 6LU7 in presence of berberine was not affected much and that why the behaviour of protein or structure of protein during simulation remain mostly unaltered. (Figure. 9).

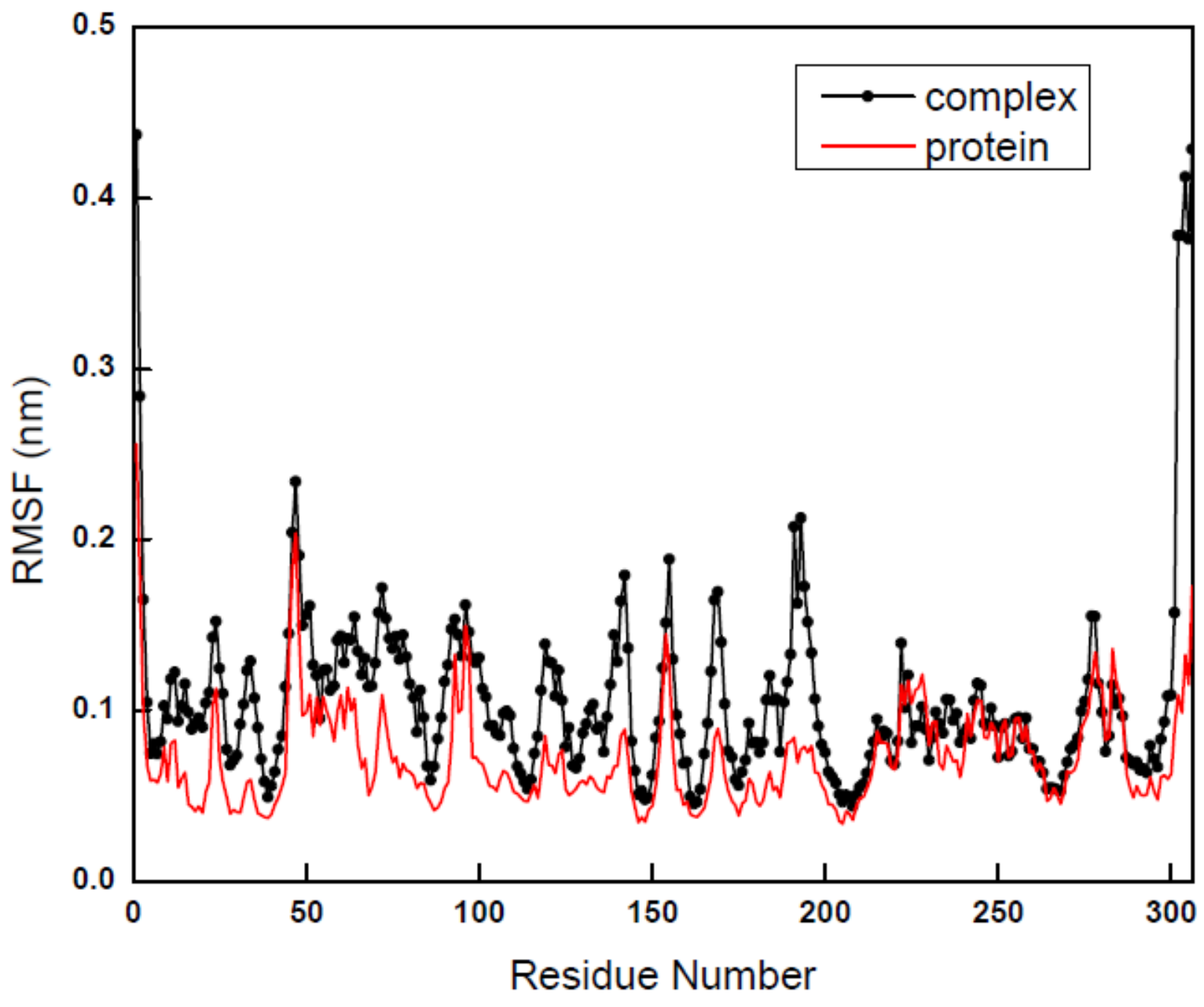

**Figure 9:** Graphs of Root mean square fluctuations (RMSF) of 6LU7 in its bare state and in berberine: 6LU7 complex.

To get an idea about the binding strength in between ligand and protein for protein: ligand complex, hydrogen bonding plays a crucial role. Ligand berberine has constant range of intermolecular hydrogen bonding with receptor protein between 0 to 2 in throughout the whole simulation process with an average value ±0.26. This result validate the nonexistence of any type of conformational change around berberine during the complexation in the binding site throughout simulation process (Figure 10). The intermolecular hydrogen bond number computed through MD simulation over the full time scale trajectory perfectly matches with the docking results also. Over all observation suggested that the berberine: 6LU7 complex is stable during simulation.

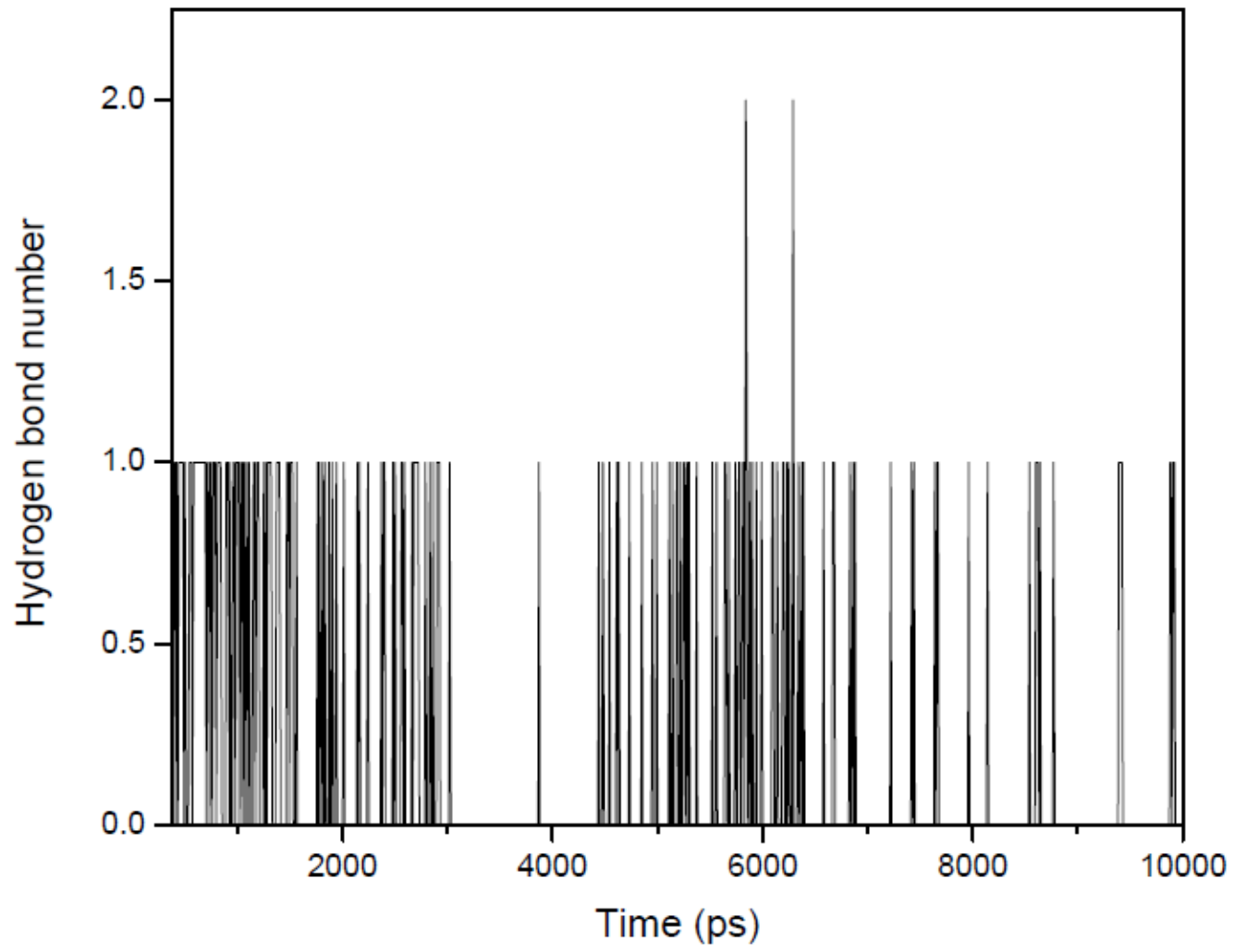

**Figure 10:** Intermolecular hydrogen bond numbers between berberine and 6LU7 in their complex (berberine: 6LU7) form for the full scale time trajectory.

The radius of gyration ($R_g$) measurement of a protein ligand complex or basic backbone protein are used to determine the compactness of the system with the time. For stably folded protein structures, $R_g$ value always like to maintain a relatively steady value for full time scale. For unfolded proteins, $R_g$ always changes over the time. Again for less compactness in the structures with conformational entropy exhibits higher $R_g$ values, while high compactness with more stability in the structure exhibits a low $R_g$ value. For the present case, $R_g$ values of both complex and base protein structures are reported in between 2.25-2.26 nm with an average value of 2.225 nm (Figure 11) which exhibited the stability of both bare protein and protein: ligand complex form. Small variation in $R_g$ values also revealed that the binding of berberine towards binding site of 6LU7 does not induce any change in the parent $CL^{pro}$ protein structure and establish the condensed stable architecture, size of the berberine: 6LU7 complex form (Fig 10). For the current paper we have shown the values over the time scale of 100 ps. The three major anti COVID-19 drugs Remdesivir Saquinavir and Darunavir also reported as showing the average $R_g$ score values as 2.2 ±0.1 nm (Khan et al. (2020)) which are exactly matching with our computed data for our proposed drug berberine.

Further to validate the applicability of berberine as proposed drug for COVID-19 we have computed the SASA of our proposed berberine: 6LU7 complex structure. SASA measures the area of receptor exposure to the solvents during the simulation process. We obtained the SASA value of receptor protein between 19 – 26 $nm^2$. The exposures of the hydrophobic part of receptor residues due to the binding of the ligand molecule towards receptor always contribute to SASA value. In the present work for both bare 6LU7 and berberine: 6LU7 complex, the computed values of SASA values were observed between 146-156 $nm^2$ which justified that ligand binding does not affect the folding of the receptor protein very much. So we may conclude that berberine: 6LU7 complex is very stable structure due to the binding of berberine with 6LU7 $CL^{pro}$ protein.

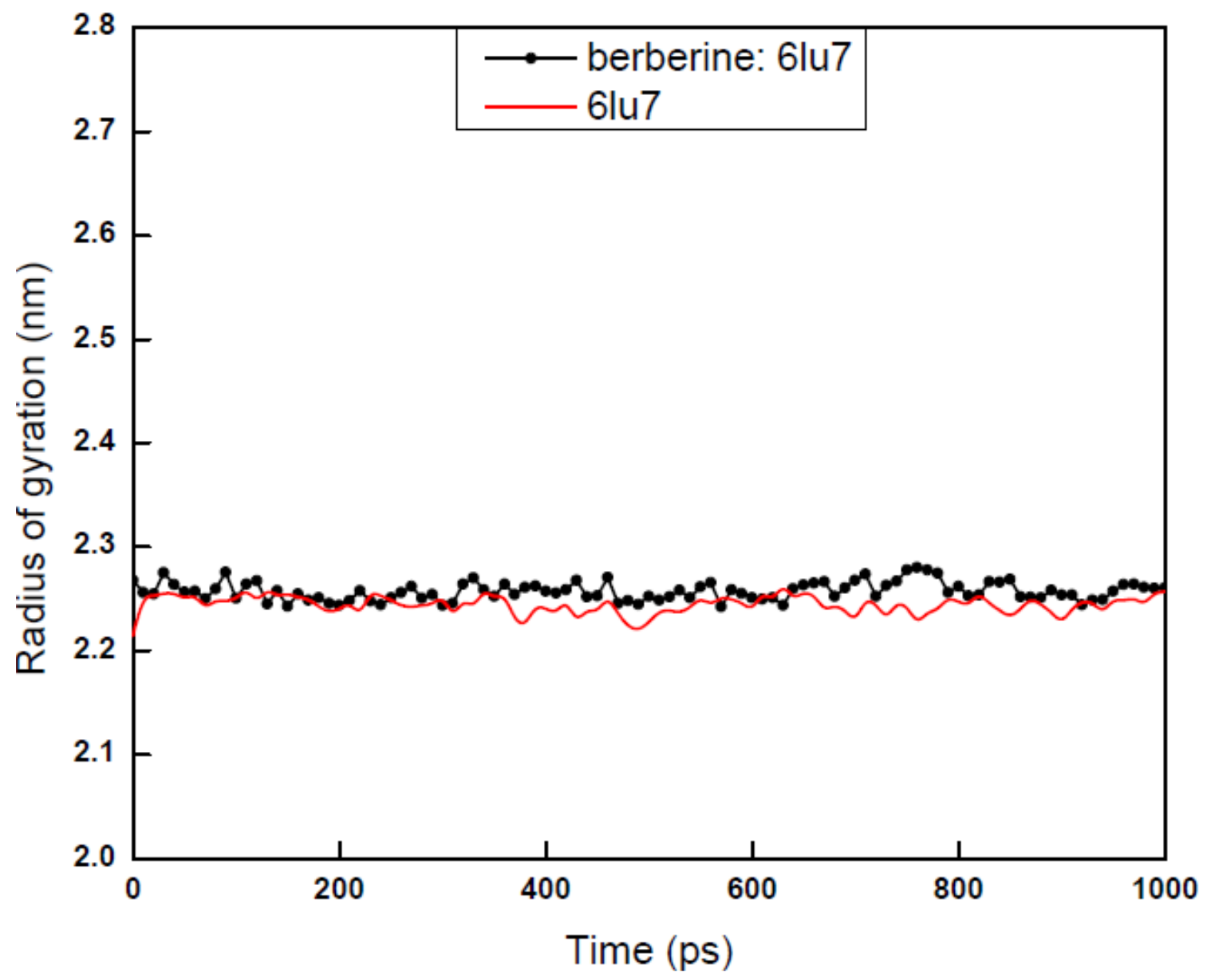

**Figure 11 :** Total radius of gyration of bare 6LU7 and berberine:6LU7 for the time trajectory 1000 ps.

To quantify the strength of the interaction between berberine and $CL^{pro}$ protease, it is always useful to compute the short-range nonbonded interaction energy between these two species which is far better than free energy or a binding energy. For protein ligand system, for free or binding energy simulation techniques, the mostly used force fields are not parametrized and so the obtained energies are in most cases become actually physically not meaningful. Whereas for nonbonded interaction energy simulation, the used force fields are parametrized to specifically target quantum mechanical interaction energies with water, so it is intrinsically balanced against meaningful quantities, and so the nonbonded interaction energy shows meaningful interpretation. For interaction energy computation we are interested in Coulombic short range protein ligand interaction energy terms and Lennard Jones short range protein ligand interaction energy terms. The total interaction energy is obtained by using these two energy values.

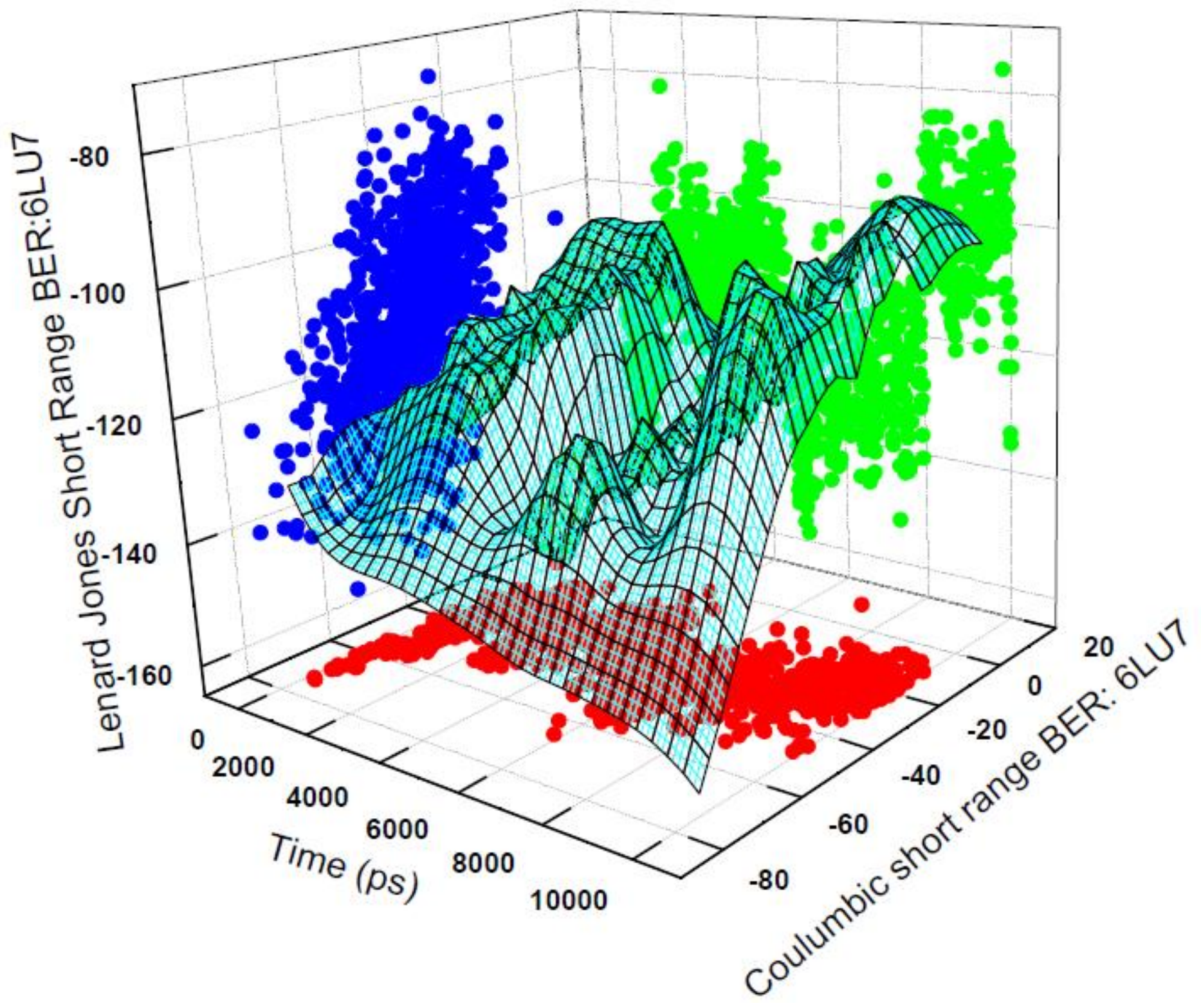

**Figure 12 :** Total nonbonded interaction energy between 6LU7 and berberine in their complex form for the time trajectory 1000 ps. Total energy is the combination of Coulombic and Lennard Jones interaction energies.

The obtained total interaction energy for the berberine: 6LU7 complex is shown by the 3D figure 12. We have obtained total interaction energy for the berberine: 6LU7 conformation in every 1000 ps time gap. The average short-range Coulombic interaction energy component compared with complex 51.7714± 3.8 kJ $mol^{-1}$ has less effect on binding affinity, whereas the average short-range Lennard-Jones energy is -130.261± 1.1kJ $mol^{-1}$ has shown strong effect on binding affinity. The total interaction energy after propagating the error according to the standard formula for addition of two quantities appeared as -186.95± 5.1kJ $mol^{-1}$. The low region of the 3D view of the time trajectory of interaction energy perfectly matches with the RMSD stable regions of the complex form and high region matches with the fluctuations of the RMSD time scale trajectory regions. All the above mentioned MD simulations data validated the good inhibition possibility of berberine towards the target $CL^{pro}$ protease 6LU7.

## 4. Conclusion:

Traditional medicinal plants are considered as a great source of input for the treatment of various diseases. For current study, we have studied the applicability of antiviral potential of some phytoconstituents extracted from a specific medicinal plant: *Tinospora cordifolia* against SARS-CoV-2 infections. *Tinospora cordifolia* is the source of various type of bioactive compounds including alkaloids, steroids, glycosides, aliphatics. Spotlight of the current study is to find an essential drug for the COVID19 disease using the antiviral activity of these compounds. Summarizing the docking and MD simulation results and interpreting the various thermodynamics parameters like temperature, pressure, density, root mean square deviation, root mean square fluctuations, inter-molecular hydrogen bonds, solvent accessible surface area, radius of gyration, potential energy, kinetic energy, total energy and interaction energies for non bonded interactions like hydrogen bonding, hydrophobic interactions, short range interactions with their analysis plots helps us to identify the best inhibitor potential drug for SERS CoV-2. The best binding docking poses of virtually screened phytochemicals berberine, β-sitosterol, octacosanol, tetrahydropalmatine, choline from the parent herb *Tinospora cordifolia* with that of the selected $3CL^{pro}$ targets I, II of main protease enzymes suggest the promising potential of these molecules to be used as raw drug material. Molecular docking and MD analysis have revealed that berberine having less binding energy and higher nonbonded interaction capability as compared to other molecules with its good binding mode of interactions. Simulation results have also revealed that berberine: $3CL^{pro}$ docked complex established better stability and can act as better inhibitor towards the CoV-2 protein compared to other inhibitors. All the reported phytochemical compounds in the present work are natural and commercially available as drugs. The present work suggests that among all the proposed phytochemicals extracted from *Tinospora cordifolia,* berberine has established its strong candidature to serve as potential inhibitors in regulating the $3CL^{pro}$ protein's function and further controlling against viral replication. Our results are expected to grab the attention of the researchers in the field of new drug discovery against SERS-CoV-2 for which till date no specific drugs or vaccines are available. The approach adopted here is general in nature and similar approach may be used to investigate the potential applications of other medicinal herbs and available drugs against COVID-19. In addition, the study may be extend to more precise investigation of protein-drug interaction using different quantum mechanical simulation methods like TDDFT, HF with modern tools like ONIOM which will help us to identify optimized and energetically favored drug:target-protein complex structures. However, such a task is computationally demanding and requires more computational resources and time. In a future communication, we would like to extend our work in that direction, so that if a clinical trial of the drug molecules studied here is aimed the same can be started with more confidence. Finally, we conclude this paper with a word caution that before using any outcome of an in silico study, proper in-vivo and in-vitro rigorous research works are to be performed.

**Acknowledgement:** Author thanks Pustak Pathak for his help in handling the simulators and software.

## References:

Anuj K, Gaurav C, Sanjeev K S, Mansi S, Pankaj T, Arvind V, Madhu S. (2020), Identification of phytochemical inhibitors againstmain protease of COVID-19 using molecularmodeling approaches. *Journal of Biomolecular Structure and Dynamics*, https://doi.org/10.1080/07391102.2020.1772112.

Anand K, Ziebuhr J, Wadhwani P, Mesters JR, Hilgenfeld R. (2003) Coronavirus main proteinase (3CLpro) structure: basis for design of anti-SARS drugs. *Science*;300 (5626):1763–7.

Beadling, C., & Slifka, M. K. (2004). How do viral infections predispose patients to bacterial infections? *Current Opinion in Infectious Diseases*, 17(3), 185–191. https://doi.org/10.1097/00001432-200406000-00003.

Bhardwaj. V J., Singh. R, Sharma. J., Rajendran. V., Purohit. R., Kumar. S., (2020). Identification of bioactive molecules from tea plant as SARS-CoV-2 main protease inhibitors., *Journal of Biomolecular Structure and Dynamics,* https://doi.org/10.1080/07391102.2020.1766572.

Boopathi. S., Poma. A. B., & Kolandaivel. P,(2020)., Novel 2019 coronavirus structure, mechanism of action, antiviral drug promises and rule out against its treatment., *Journal of Biomolecular Structure and Dynamics*, https://doi.org/10.1080/07391102.2020.1758788.

Berendsen, H.J.C., van der Spoel, D., & van Drunen, R. (1995). GROMACS: a messagepassing parallel molecular dynamics implementation. Computer *PhysicsCommunications, 91*(1-3), 43-56.

Becke A. (1997) A new inhomogeneity parameter in density-functional theory,*J. Chem. Phys*. 107, 8554. https://doi.org/10.1063/1.476722.

Chang, C., Lo, S.-C., Wang, Y.-S., & Hou, M.-H. (2016). Recent insights into the development of therapeutics against coronavirus diseases by targeting N protein. *Drug Discovery Today*, *21*(4), 562–572.

Corbin KD, Zeisel SH, (2012) "Choline metabolism provides novel insights into nonalcoholic fatty liver disease and its progression". *Current Opinion in Gastroenterology*. . 28 (2): 159–65.

de Wit, van Doremalen, N., Falzarano, D., & Munster, V. J. (2016). SARS and MERS: Recent insights into emerging coronaviruses. *Nature Reviews Microbiology*, *14*(8), 523–534.

Doremalen N, Morris. D, Bushmaker. T . (2020), Aerosol and Surface Stability of SARS-CoV-2 as compared with SARS-CoV-1. *New Engl J Med* https://doi.org/10.1056/NEJMc2004973.

Dayer MR, Taleb-Gassabi S, Dayer MS. Lopinavir; (2017) a potent drug against coronavirus infection: insight from molecular docking study. *Arch Clin Infect Diseas* ;12:e13823.

*Dassault Systèmes BIOVIA, Discovery Studio Modeling Environment, Release 2017, San Diego: Dassault Systèmes,.*

Daina A, Michielin O, Zoete V (2017) SwissADME: a free web tool to evaluate pharmacokinetics, drug-likeness and medicinal chemistry friendliness of small molecules. Sci Rep 7:42717. https://doi.org/10.1038/srep42717.

Dewick, P. (2009) Medicinal Natural Products: *A Biosynthetic Approach (3rd ed.).* West Sussex, England: Wiley. p. 357. ISBN 978-0-471-49641-0.

Enmozhi, S.K., Raja, K., Sebastine, I., & Joseph, J. (2020). Andrographolide as a potential inhibitor of SARS-CoV-2 main protease: an in-silico approach. *Journal of Biomolecular Structure and Dynamics*.

Feixiong C, (2013) Adverse Drug Events: Database Construction and in Silico Prediction, *J. Chem. Inf. Model.*, 53, 4, 744–752.

Fehr, A. R., & Perlman, S. (2015). Coronaviruses: An overview of their replication and pathogenesis. *Methods in Molecular Biology*, 1282, 1–23. 10.1007/978-1-4939-2438-7_1.

Frisch, M, J., (2004) Gaussian 09, revision D.01; Gaussian Inc.: Wallingford CT.

Gorden, D.E., Gwendolyn, M.J., Bouhaddou, M., et al. (2020). A SARS-CoV-2-human protein-protein interaction map reveals drug targets and potential drug-repurposing. *bioRxiv*. https://doi.org/10.1101/2020.03.22.002386.

Guo, Y., Cao, Q., Hong, Z., Tan, Y., Chen, S., Jin, H., Tan, K., Wang, D., & Yan, Y. (2020). The origin, transmission and clinical therapies on coronavirus disease 2019 (COVID-19) outbreak – an update on the status. *Military Medical Research*, 7(1), 11. https://doi.org/10.1186/s40779-020-00240-0.

Guan, W. J., Ni, Z. Y., Hu, Y., Liang, W. H., Ou, C. Q., He, J. X., Liu, L., Shan, H., Lei, C. L., Hui, D., Du, B., Li, L. J., Zeng, G., Yuen, K. Y., Chen, R. C., Tang, C. L., Wang, T., Chen, P. Y., Xiang, J., Li, S. Y, . China Medical Treatment Expert Group for Covid-19 (2020). Clinical characteristics of Coronavirus disease 2019 in China. *The New England Journal of Medicine*, NEJMoa2002032. Advance online publication. https://doi.org/10.1056/NEJMoa2002032.

Gorbalenya, A. E., Baker, S. C., Baric, R. S., de Groot, R. J., Drosten, C., Gulyaeva, A. A., Haagmans, B. L., Lauber, C., Leontovich, A. M., Neuman, B. W., Penzar, D., Perlman, S., Poon, L. L. M., Samborskiy, D. V., Sidorov, I. A., Sola, I., & Ziebuhr, J. (2020). The species Severe acute respiratory syndrome-related coronavirus: Classifying 2019-nCoV and naming it SARS-CoV-2. *Nature Microbiology*, https://doi.org/10.1038/s41564-020-0695-z.

Graham, R. L., Donaldson, E. F., & Baric, R. S. (2013). A decade after SARS: Strategies for controlling emerging coronaviruses. *Nature Reviews. Microbiology*, 11(12), 836–848. https://doi.org/10.1038/nrmicro3143.

Grein, J., Ohmagari, N., Shin, D., Diaz, G., Asperges, E., Castagna, A., & Feldt, T. (2020). Compassionate use of remdesivir for patients with severe Covid-19. *New England Journal of Medicine*, 1–10. https://doi.org/10.1056/NEJMoa2007016.

Gao, Y., Gao, Y., Yan, L., Huang, Y., Liu, F., Zhao, Y., Cao, L., Tao, W., Sun, Q., Ming, Z., Zhang, L., Ge, J., Zheng, L., Zhang, Y., Wang, H., Zhu, Y., Zhu, C., Hu, T., Hua, T., et al. (2020, April). Structure of the RNAdependent RNA polymerase from COVID-19 virus. *Science*, 7498, 1–9.

Giacomini KM, Huang SM, Tweedie DJ, Benet LZ, Brouwer KL, Chu X, Dahlin A, Evers R, Fischer V, Hillgren KM, Hoffmaster KA, Ishikawa T, Keppler D, Kim RB, Lee CA, Niemi M, Polli JW, Sugiyama Y, Swaan PW, Ware JA, Wright SH, Wah Yee S, Zamek-Gliszczynski MJ, Zhang L. (2010) The International Transporter Consortium. Membrane transporters in drug development. Nat. Rev. Drug Discov. ;9:215–236.

Gupta, M. K., Vemula, S., Donde, R., Gouda, G., Behera, L., & Vadde, R. (2020). In-silico approaches to detect inhibitors of the human severeacute respiratory syndrome coronavirus envelope protein ion channel. *Journal of Biomolecular Structure and Dynamics*, 1–17. https://doi.org/10.1080/07391102.2020.1751300.

Gajula, M.N.P., Kumar, A., & Ijaq, J. (2016). Protocol for molecular dynamics simulations of proteins. *Bio-protocol, 6,* 1-11. doi:10.21769/BioProtoc.2051.

Hendaus M A,. (2020), Remdesivir in the treatment of coronavirus disease 2019 (COVID-19): a simplified summary, *Journal of Biomolecular Structure and Dynamics*, https://doi.org/10.1080/07391102.2020.1767691.

Jin, Z., Du, X., Xu, Y., Deng, Y., Liu, M., Zhao, Y., Zhang, B., Li, X., Zhang, L., Peng, C., Duan, Y., Yu, J., Wang, L., Yang, K., Liu, F., Jiang, R., Yang, X., You, T., Liu, X., … Yang, H. (2020). Structure of Mpro from COVID-19 virus and discovery of its inhibitors. Nature. https://doi.org/10.1038/s41586-020-2223-y

Jee, B., Kumar, S., Yadav, R., Singh, Y., Kumar, A., & Sharma, M. (2018). Ursolic acid and carvacrol may be potential inhibitors of dormancy protein small heat shock protein16.3 of *Mycobacterium tuberculosis*, *Journal of Biolomolecular Structure and Dynamics, 36*(13), 3434-3443.

Jiang, F., Deng, L., Zhang, L., Cai, Y., Cheung, C. W., & Xia, Z. (2020,March 4). Review of the clinical characteristics of coronavirus disease 2019 (COVID-19). *Journal of General Internal Medicine*. 10.1007/s11606-020-05762-w.

Khan, S. A., Zia, K., Ashraf, S., Uddin, R., & Ul-Haq, Z. (2020). Identification of chymotrypsin-like protease inhibitors of SARS-Cov-2 via integrated computational approach. *Journal of Biomolecular Structure and Dynamics,* 1–10. https://doi.org/10.1080/07391102.2020.1751298.

Kaisari, E., & Borruat, F. (2020). Keeping an eye on hydroxychloroquine retinopathy. *KlinischeMonatsblätterfürAugenheilkunde*. Efirst.

Kumar, A., Kumar, R., Sharma, M., Kumar, U., Gajula, M.NV.P., & Singh, K.P. (2018). Uttarakhand medicinal plants database (UMPDB): a platform for exploring genomic, chemical, and traditional knowledge. *Data (MDPI), 3*(1), 7. doi: 10.3390/data3010007.

Khaerunnisa, S., Kurniawan, H., Awaluddin, R., Suhartati, S., & Soetjipto, S. (2020). Potential inhibitor of COVID–19 main protease ($M_{pro}$) from several medicinal plant compounds by molecular docking study. *Preprints,* 2020030226.

Lipinski CA. (2004) "Lead- and drug-like compounds: the rule-of-five revolution". *Drug Discovery Today: Technologies.* 1 (4): 337–341.

Mohamed A. Hendaus, Fatima A. Jomha., (2020). Covid-19 induced superimposed bacterial infection. *Journal of Biomolecular Structure and Dynamics,* https://doi.org/10.1080/07391102.2020.1772110.

Mesecar, A.D. (2020) A taxonomically-driven approach to development of potent, broad-spectrum inhibitors of coronavirus main protease including SARS-CoV-2 (COVID-19). https://www.rcsb.org/structure/6W63. DOI: 10.2210/pdb6W63/pdb.

Michel F. S. (1999) Python: A Programming Language for Software Integration and Development. *J. Mol. Graphics Mod.*, , Vol 17, February. pp57-61.

Mayo S.L, Olafson, B, D, Goddard, W, A. , (1990) a generic force field for molecular Simulations, . *Phys. Chem.*, 94, 26, 8897–8909.

Patil1. R., Das. S, Stanley. A., Yadav. L., Sudhakar. A., Varma. A.K., (2010), Optimized Hydrophobic Interactions and Hydrogen Bonding at the Target-Ligand Interface e Pathways of Drug-Designing, *PLoS One.* 5(8): e12029. Issue 8 | e12029.

Rothe, C., Schunk, M., Sothmann, P., Bretzel, G., Froeschl, G., Wallrauch, C., Zimmer, T., Thiel, V., Janke, C., Guggemos, W., Seilmaier, M., Drosten, C., Vollmar, P., Zwirglmaier, K., Zange, S., W€olfel, R., & Hoelscher, M. (2020). Transmission of 2019-nCoV infection from an asymptomatic contact in Germany. *New England Journal of Medicine*, 382(10), 970–971. https://doi.org/10.1056/NEJMc2001468.

Rameez J K., Rajat K J., Gizachew M A., Monika J, Ekampreet S., Amita P., Rashmi P S, Jayaraman M., & Amit K S.,(2020) Targeting SARS-CoV-2: a systematic drug repurposing approach to identify promising inhibitors against 3C-like proteinase and 2′-Oribose methyltransferase. *Journal of Biomolecular Structure and Dynamics*, https://doi.org/10.1080/07391102.2020.1753577.

Rana, M., & Chowdhury, P. (2017) Perturbation of hydrogen bonding in hydrated pyrrole-2-carboxaldehyde complexes. *J. Molecular Modelling (Springer),* 23, 216.

Rudkowska I, AbuMweis SS, Nicolle C, Jones PJ. (2008) "Cholesterol-lowering efficacy of plant sterols in low-fat yogurt consumed as a snack or with a meal*". J Am Coll Nutr..* 27 (5): 588–95.

Salata, C., Calistri, A., Parolin, C., & Palù, G. (2019). Coronaviruses: A paradigm of new emerging zoonotic diseases. *Pathogens and Disease*, *77*(9), ftaa006.

Sonkamble. V. V., Kamble. L.H., (2015) Antidiabetic potential and identification of phytochemicals from Tinospora cordifolia, *Am. J. Phytomed.* Clin. Ther. 3 (2015) 97–110.

Sharma. P., , Dwivedee. B.P., , Bisht. D., , Dash. A.K.,, Kumar. D., (2019). The chemical constituents and diverse pharmacological importance of Tinospora cordifolia, *Heliyon*. 5 e02437, https://doi.org/10.1016/j.heliyon.2019.e024.

Singla, N., & Chowdhury, P. (2014) Inclusion behaviour of Indole-7-Carboxaldehyde inside β-cyclodextrin: A nano cage. *Chemical Physics., 441*, 93-100.

Sarma, P., Sekhar, N., Prajapat, M., Avti, P., Kaur, H., Kumar, S., Singh, S., Kumar, H., Prakash, A., Dhibar, D.P., Medhi, B. (2020). In-silico homology assisted identification of inhibitor of RNA binding against 2019- nCoV N-protein (N terminal domain). *Journal of Biomolecular Structure& Dynamics*, 1–11. https://doi.org/10.1080/07391102.2020.1753580.

Soteras Gutierrez, I., Lin, F.-Y., Vanommeslaeghe, K., Lemkul, J.A., Armacost, K.A., Brooks, Cl., III, and MacKerell, A.D., Jr., (2016) "Parametrization of Halogen Bonds in the CHARMM General Force Field: Improved treatment of ligand-protein interactions," *Bioorganic & Medicinal Chemistry*, https://doi.org/10.1016/j.bmc.2016.06.034.

Trott, O., & Olson, A. J. (2010),.AutoDock Vina: improving the speed and accuracy of docking with a new scoring function, efficient optimization, and multithreading. *J Comput Chem*, 31(2), 455-461. doi:10.1002/jcc.21334.

Taylor, J. C; (2003) Rapport, Lisa; Lockwood, G.Brian "Octacosanol in human health". *Nutrition.*. 19 (2): 192–5.

van der Hoek, L., Pyrc, K., Jebbink, M. F., Vermeulen-Oost, W., Berkhout, R. J., Wolthers, K. C., Wertheim-van Dillen, P. M., Kaandorp, J., Spaargaren, J., & Berkhout, B. (2004). Identification of a new human coronavirus. *Nature Medicine*, 10(4), 368–373. https://doi.org/10.1038/nm1024.

Velavan, T. P., & Meyer, C. G. (2020). The COVID-19 epidemic. *Tropical Medicine & International Health: TM & IH*, 25(3), 278–280. https://doi.org/10.1111/tmi.13383.

Veber, D. F., Johnson, S. R., Cheng, H. Y., Smith, B. R., Ward, K. W., & Kopple, K. D..(2002) Molecular properties that influence the oral bioavailability of drug candidates. *Journal of medicinal chemistry*, 45(12), 2615-2623.

Vanommeslaeghe, K., and MacKerell Jr., A.D., (2012) "Automation of the CHARMM General Force Field (CGenFF) I: bond perception and atom typing," *Journal of Chemical Informationa and Modeling*, 52: 3144-3154, , PMC3528824.

Wang, D., Hu, B., Hu, C., Zhu, F., Liu, X., Zhang, J., Wang, B., Xiang, H., Cheng, Z., Xiong, Y., Zhao, Y., Li, Y., Wang, X., & Peng, Z. (2020). Clinical characteristics of 138 hospitalized patients with 2019 novel coronavirus–infected pneumonia in Wuhan, China. *JAMA*, 323(11), 1061. https://doi.org/10.1001/jama.2020.1585.

Weiss, P., & Murdoch, D.R. (2020). Clinical course and mortality risk of severe COVID-19.*The Lancet, 395*(10229), 1014-1015.

Wang, M., Cao, R., Zhang, L., Yang, X., Liu, J., Xu, M., & Xiao, G. (2020). Remdesivir and chloroquine effectively inhibit the recently emerged novel coronavirus (2019- nCoV) *in vitro. Cell Research, 30*(3), 269–271.

Wang, B., Guo, H., Ling, L., Ji, J., Niu, J., & Gu, Y. (2020). The chronic adverse effect of chloroquine on kidney in rats through an autophagy dependent and independent pathways. *Nephron, 144*(1), 53–64.

Woo, P. C. Y., Lau, S. K. P., Chu, C-m., Chan, K-h., Tsoi, H-w., Huang, Y., Wong, B. H. L., Poon, R. W. S., Cai, J. J., Luk, W-k., Poon, L. L. M., Wong, S. S. Y., Guan, Y., Peiris, J. S. M., & Yuen, K-Y. (2005). Characterization and complete genome sequence of a novel coronavirus, coronavirus HKU1, from patients with pneumonia. *Journal of Virology*, *79*(2), 884–895.

Wu, F., Zhao, S., Yu, B., Chen, Y.-M., Wang, W., Song, Z.-G., Hu, Y., Tao, Z.- W., Tian, J.-H., Pei, Y.-Y., Yuan, M.-L., Zhang, Y.-L., Dai, F.-H., Liu, Y., Wang, Q.-M., Zheng, J.-J., Xu, L., Holmes, E. C., & Zhang, Y.-Z. (2020 March). A new coronavirus associated with human respiratory disease in China. *Nature*, 579(7798), 265–269. doi:10.1038/s41586-020-2008-3.

Wang Z, Guo X, Liu Z, Cui M, Song F, Liu S: (2008) Studies on alkaloids binding to GC-rich human survivin promoter DNA using positive and negative ion electrospray ionization mass spectrometry. *J Mass Spectrom*. Mar;43(3):327-35.

Yuriev, E.; Agostino, M.; Ramsland, P.A. (2011) Challenges and advances in computational docking: 2009 in review. *J. Mol. Recognit*., 24, 149–164.

Zhang, L., Lin, D., Sun, X., Curth, U., Drosten, C., Sauerhering, L., Becker, S., Rox, K., & Hilgenfeld, R. (2020). Crystal structure of SARS-Cov-2 main protease provides a basis for design of improved a-ketoamide inhibitors. Science (New York, N.Y), 368(6489), 409–412. https://doi.org/10.1126/science.abb3405.